\documentclass[preprint,authoryear,12pt]{elsarticle}

\usepackage{epsfig}
\usepackage{amssymb}

\usepackage[
a4paper=true,%
breaklinks=true,%
colorlinks=true,%
pdfauthor={First Author et al.},%
pdftitle={Template for manuscripts in Advances in Space Research}%
]{hyperref}

\journal{Advances in Space Research}

\begin{document}

\begin{frontmatter}




\title{Observing the Energetic Universe at Very High Energies with the VERITAS Gamma Ray Observatory}


\author{R. Mukherjee for the VERITAS Collaboration}

\address{Department of Physics and Astronomy, Barnard College, Columbia University, New York, NY 10027, USA}

\begin{abstract}
Very high energy gamma-ray observations offer indirect methods for studying the highest energy cosmic rays in our Universe. The origin of cosmic rays at energies greater than $10^{18}$ eV remains a mystery, and many questions in particle astrophysics exist.  The VERITAS observatory in southern Arizona, USA, carries out an extensive observation program of the gamma-ray sky at energies above $85$ GeV. Observations of Galactic and extragalactic sources in the TeV band provide clues to the highly energetic processes occurring in these objects, and could provide indirect evidence for the origin of cosmic rays and the sites of particle acceleration in the Universe. VERITAS has now been operational for ten years with the complete array of four atmospheric Cherenkov telescopes. In this review, we present the status of VERITAS, and give few results from three of its key scientific programs: extragalactic science, Galactic physics, and study of fundamental physics and cosmology. 

 \end{abstract}


\end{frontmatter}


\section{Introduction}
\label{}

VERITAS (Very Energetic Radiation Imaging Telescope Array System) is an array of four imaging atmospheric Cherenkov telescopes (IACTs), located at the base camp of the Fred Lawrence Whipple Observatory (FLWO) in southern Arizona. VERITAS uses ground-based detection techniques to explore the Universe at very high energy (VHE) gamma rays from 85 GeV to 30 TeV. High energy gamma-rays from astrophysical sources emitted in the direction of Earth, interact with the Earth's atmosphere,  producing electromagnetic cascades which generate short-lived but bright flashes of Cherenkov radiation, at a characteristic angle. IACTs are capable of detecting these brief flashes of Cherenkov light in the visible and UV range, and thus indirectly function as gamma-ray telescopes. However, at a much higher rate than gamma-ray showers, cosmic rays also produce showers of charged particles and Cherenkov radiation, which constitute for the principal source of background for gamma-ray telescopes.  IACTs such as VERITAS utilize the Cherenkov light emitted from air showers to form an image on the camera-plane of the longitudinal and lateral development of the air shower. 
The imaging technique can reject 99.7\% of the cosmic-ray background using statistical analysis methods, while retaining $> 50$\% of the gamma-ray signal. Shower images can be modeled as an ellipse whose semi-major axis points back to the origin of the shower. Second-moment parameters, or  Hillas parameters~\citep{hillas}, characterizing the length, width and asymmetry of the image are calculated and used to discriminate between primary particle types. 
By employing multiple telescopes in combination (four in the case of VERITAS) to increase the collection area and obtain a stereoscopic image of the particle showers in order to reduce cosmic ray background, VERITAS operates at a sensitivity an order of magnitude better than the previous generation of telescopes~\citep{horan_weekes2003}. At the present time, in addition to VERITAS, the other major IACTs in operation are MAGIC~\citep{magic_review} and H.E.S.S.~\citep{hess_review}. FACT (First G-APD Cherenkov Telescope), a new telescope using advanced technology in the form of a camera comprised of Geiger avalanche photo diodes, has been successfully monitoring blazars at TeV energies since 2011~\citep{fact_review}.

\begin{figure}[t!]
\begin{center}
\includegraphics[scale=0.54]{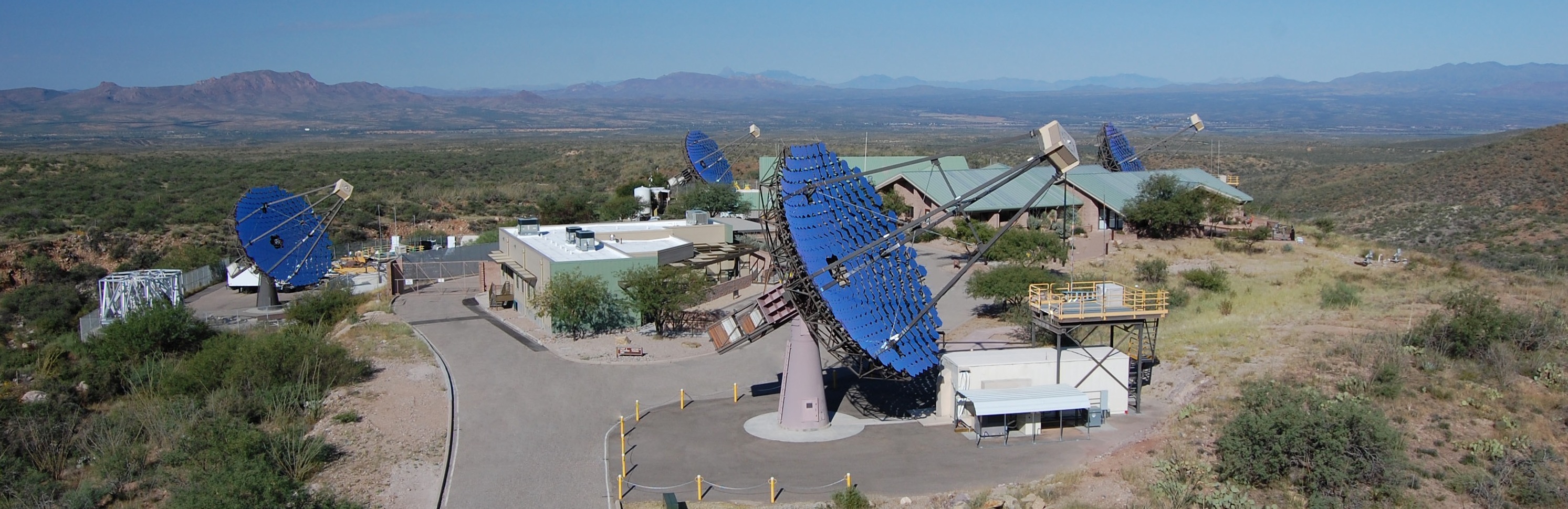} 
\end{center}
\caption{A photograph of the VERITAS telescope array, located at the base camp of the Whipple Observatory in southern Arizona. This configuration of the array has been in use since 2009. The building at the center of the photograph is the VERITAS control room. Source: Steve Criswell and the VERITAS website~\citep{veritas_spec}.}
\end{figure}

The VERITAS telescopes are of Davies-Cotton design, each with a diameter of 12 m. Each telescope has  a camera at its focal plane comprising 499 photomultiplier tubes (PMTs) arranged in a hexagonal pattern with a field of view of diameter 3.5$^\circ$. Figure 1 is a photograph of the VERITAS observatory, showing the array of four telescopes. Figure 2 shows a detail of the VERITAS camera focal plane and an image of a gamma-ray event recorded by all four telescopes. Further details of the VERITAS telescopes including image analysis, telescope specifications, background rejection techniques and scientific analysis of data may be found in {\it VERITAS: Status and Highlights}~\citep{veritas_details}. 

\begin{figure}[t!]
\begin{center}
\includegraphics[scale=0.4]{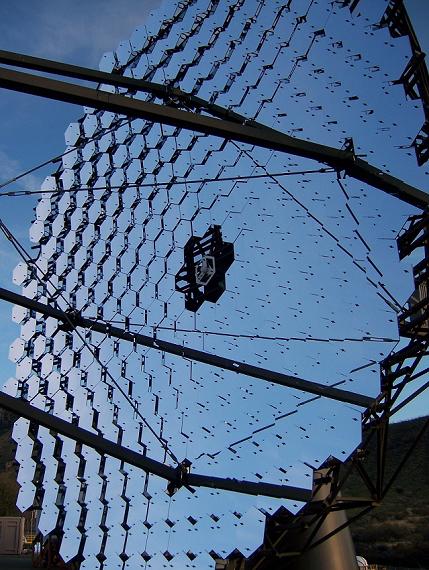} \hfill
\includegraphics[scale=1.3]{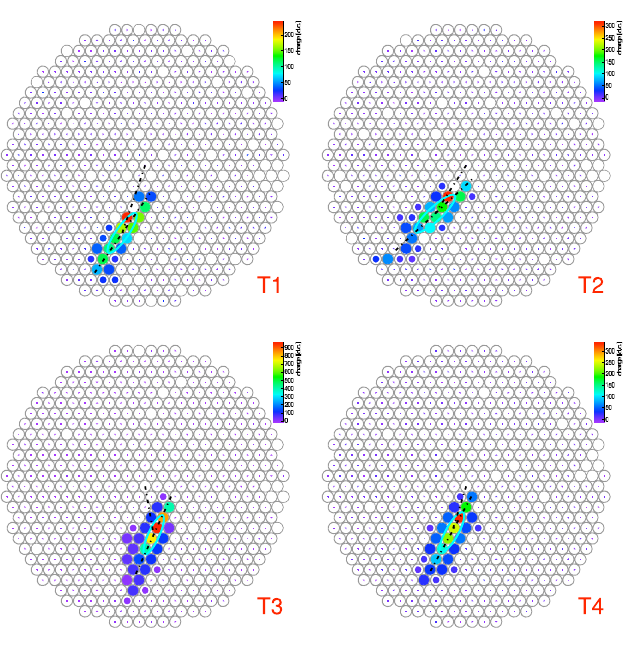} 
\end{center}
\caption{(a) Left: A VERITAS telescope showing the facet structure of the reflector. The telescope diameter is roughly 12 m. (b) Right: Camera view of a 4-telescope event that is likely to be a gamma-ray event. The color scale indicates the integrated charge per channel, and ranges from 0 (purple) to 900 (red), and goes in steps of 100 digital counts (d.c.). Each camera has a total of 499 pixels. (Source: VERITAS website.) 
}
\end{figure}

VERITAS has been operating with the complete four-telescope array since 2007. In the past decade, there have been several upgrades and improvements to VERITAS, such as the relocation of  one of the telescopes in order to obtain a more symmetric array layout in 2009~\citep{Kieda2013}, implementation of a new alignment system which improved the instrument point spread
function~\citep{McCann2010} and allowed for a 30\% improvement in sensitivity, and a complete upgrade of the VERITAS cameras with new photomultiplier tubes in the summer of 2012~\citep{Otte2011}. These new PMTs, with super bialkali photocathode material, allowed a peak quantum efficiency of greater than 32\%, leading to an increase of the photon detection efficiency of each camera by approximately 50\%. VERITAS is now able to detect a source with a Crab Nebula-like spectrum and a flux of 1\% Crab Nebula strength in about 25 hours in less than half of the exposure required for the original array configuration from 2007. VERITAS has  an angular resolution of less than $0.1^\circ$ (68\% containment) at 1 TeV and an energy resolution of 15\% to 20\%.  Figure 3 shows the differential sensitivity of VERITAS for the three different sets of ``cuts"~\citep{npark2015}. Soft cuts provide the highest sensitivity at approximately 100 GeV, while the hard cuts are optimized for energies greater than 600 GeV. VERITAS now typically collects approximately  1300 hours of data per year, including 200 hours taken in moderate moonlight with the lunar disk up to 50\% illuminated. 

\begin{figure}[b!]
\begin{center}
\includegraphics[scale=0.4]{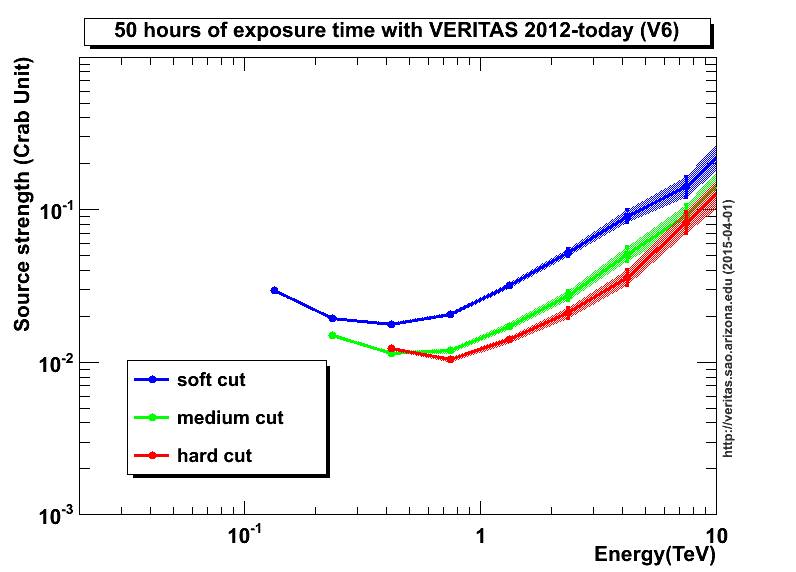} 
\end{center}
\caption{ VERITAS current differential sensitivity estimated by using Crab Nebula data. Differential sensitivities give the strength of a source VERITAS can detect with a significance of 5 sigma, assuming an exposure time of 50 hours. The source is assumed to be at elevations above 70 degrees~\citep{npark2015}. The latest VERITAS sensitivity curves are available on the VERITAS website \citep{veritas_spec}.}
\end{figure}

The scientific program of VERITAS includes the study of extragalactic sources such as blazars, galactic systems such as pulsars,  supernova remnants, pulsar wind nebulae, binary systems, and studies of fundamental physics and  cosmology. Together with 
the Fermi Large Area Telescope ({\sl Fermi}-LAT)~\citep{fermilat}, the High Altitude Water Cherenkov (HAWC) Experiment~\citep{Abeysekara2017}, and the IceCube neutrino observatory~\citep{icecube}, VERITAS offers the chance to use neutral messengers to directly probe cosmic accelerators. 
With the detection of astrophysical neutrinos by IceCube~\citep{aartsen2013} and reports of the first gravitational wave localizations by LIGO~\citep{ligo_prl2016} and VIRGO~\citep{virgo}, VERITAS has the potential for multi-messenger complementarity, and could detect electromagnetic counterparts for observations made by LIGO/VIRGO or IceCube.  The first-ever detections of electromagnetic counterparts to a gravitational wave was recently made for the LIGO event 20170817~\citep{ligo170817} by {\sl Fermi}-GBM and INTEGRAL~\citep{GBM170817}, demonstrating the promise of multi-messenger astronomy. With its superior angular resolution, and its ability to follow up on transient triggers alerts, VERITAS therefore plays a critical role in multi-messenger astrophysics. 

In the past three years VERITAS has been functioning as a robust observatory, with a steady observing program, and the scientific output of VERITAS has remained strong.  The VERITAS source catalog, as of this publication, is shown in Fig. 4.  The source count stands at 59 sources and is composed of eight different source classes.  As of this writing, a comprehensive set of VERITAS results were presented at the 35th International Cosmic Ray Conference (ICRC)~\citep{icrc2017proc}. In the following we present a very brief summary of a few selected results on individual sources. We note that the field of very high energy astrophysics has been enriched with the results from all three major IACTs, H.E.S.S., MAGIC and VERITAS. However, in this review we focus just on the VERITAS results.

\begin{figure}[b!]
\begin{center}
\includegraphics[scale=0.50]{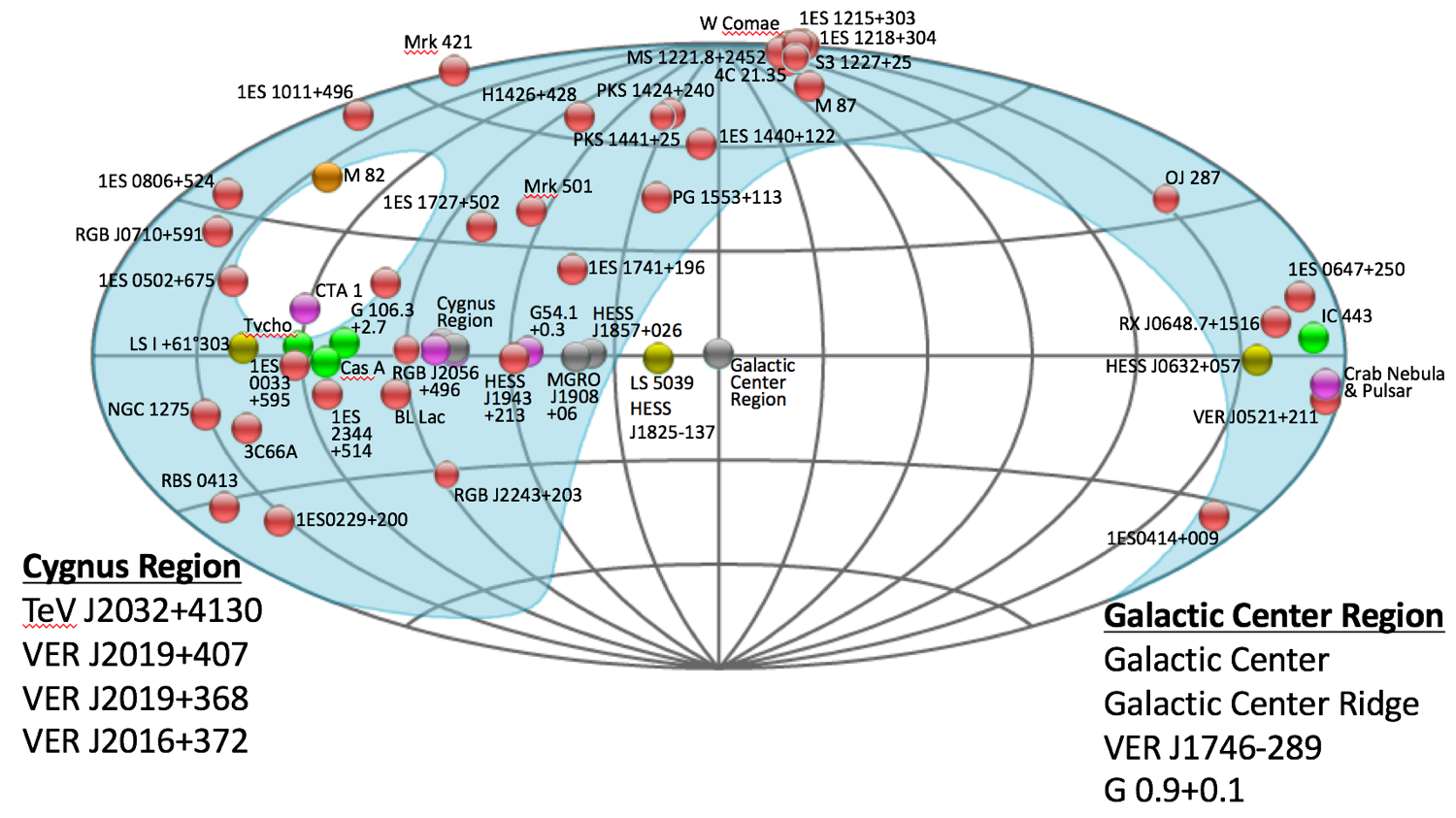} 
\end{center}
\caption{VERITAS skymap (October 2017). The shaded area shows the region of visibility for VERITAS, which is located in the Northern Hemisphere. Different source types are represented as follows: blazars (red), pulsar wind nebulae (magenta), binaries (yellow), supernova shells (green), starburst galaxies (orange), and dark/unidentified (grey). Figure modified from TeVCat \citep{tevcat}. }
\end{figure}

\section{Extragalactic Source Studies with VERITAS} 

The majority of the sources detected by VERITAS, as seen in Fig. 4, are active galactic nuclei (AGN) of the blazar class, with gamma-ray emission originating in their relativistic jets. TeV observations of AGN help us constrain models of particle acceleration and energy dissipation in blazar jets, and help us determine the size and location of the gamma-ray emission region. 
Blazars are also important probes of the extragalactic background light (EBL), which absorbs their TeV photons traveling cosmological distances.
Furthermore, studies of blazars at the highest energies may  be used to explore and constrain the weak intergalactic magnetic fields (IGMF), and test the validity of the Lorentz Invariance principle at high energies. Blazars observations constitute a key component of the VERITAS observation program. Over the past decade the VERITAS observatory has carried out a  combination of dedicated blazar observations, long term studies of certain historical blazars, a discovery program for detecting new TeV blazars, and target-of-opportunity observations for flaring sources~\citep{wystan_icrc2017}. 

Table 1 lists all the extragalactic sources detected by VERITAS as of October 2017. Except for the two FR I type radio galaxies, M87 and NGC 1275, and the starburst galaxy M82, all the extragalactic sources are active galaxies of the blazar class. It is not surprising that of the 36 sources shown in the table, 23 are high-frequency-peaked BL Lac objects (HBLs). As we have learnt, the TeV blazar population is largely dominated by high-frequency peaked BL Lacs, in which the peaks in the spectral energy distribution (SED) lie in the X-ray and TeV bands~\citep{tevcat}. 
These blazars exhibit strongly correlated TeV and X-ray emission on many occasions, suggesting that the low-energy peak is explained by synchrotron emission from a population of ultra relativistic electrons, and that the high-energy peak arises from inverse-Compton scattering of the synchrotron photons by the same population of electrons. So called ``SSC," or 1-zone synchrotron self-Compton models of blazar emission~\citep{tevssc}, have been found to be largely successful in explaining TeV emission in blazars. In addition to HBLs, VERITAS has also detected high energy emission from other blazar classes, such as intermediate- and low-frequency-peaked BL Lacs (IBLs/LBLs), with gamma-ray emission peaking at lower frequencies, flat spectrum radio quasars (FSRQs), and radio galaxies. Typically, FSRQs are only detected during flaring episodes. As the catalog of TeV LBLs , IBLs and FSRQs grows, it will be be possible to carry out studies of blazar populations, and the study of jet properties in systems with different supermassive black hole (SMBH) masses and accretion rates than the commonly observed HBLs. 

The blazars detected by VERITAS are predominantly nearby, with the range of redshifts from 0.03 to at least 0.939. It is interesting to note that all VERITAS AGN are also detected by the {\sl Fermi}-LAT at lower energies. This could possibly be due to observational bias, since the strategy of the VERITAS observations has been to follow up on hard-spectrum {\sl Fermi}-LAT sources, or flaring {\sl Fermi}-LAT blazars, as well as X-ray selected HBLs, which have also been detected at GeV energies. Only a handful of TeV detected blazars are not seen at {\sl Fermi}-LAT energies. One example is the HBL 1ES 0347-121 \citep{tevcat}, and some studies have used this non-detection to place interesting limits on the IGMF~\citep{vovk2010}. VERITAS observations are supported by simultaneous multi-wavelength data that has helped in modeling studies of the blazar SEDs. 

\subsection{Flat-Spectrum Radio Quasars and the Distant Blazar PKS 1441+25} 

There are only 6 FSRQs known to be TeV emitters as of date~\citep{tevcat}, and only two detected by VERITAS, namely, PKS 1441+25 and PKS 1222+216. At a redshift of 0.939, PKS 1441+25 is the most distant source in the VERITAS catalog. FSRQs are believed to host radiatively efficient disks that enrich the environment of the supermassive black hole with ultraviolet-to-optical photons. In FSRQs, $\gamma\gamma$ interaction of TeV photons with the ultraviolet-to-optical photon field from the disk, the reprocessed emission from the clouds of the broad line region (BLR), or the infrared radiation (IR) from the ``dusty torus'' can make the environment opaque to TeV gamma rays from the base of the jet~\citep{donea2002}. VERITAS detected gamma-ray emission from PKS 1441+25 from about 80 GeV up to 200 GeV in 2015 April, a period when the source was highly active across all wavelengths. The observation results are summarized in the {\it Astrophysical Journal Letters}~\citep{veritas1441_25}, which was simultaneously published with the MAGIC detection of the source at the same time~\citep{magic_1441}. 

Figure 5a shows the broadband SED of PKS 1441+25 from X-ray to TeV gamma rays. A stationary model was used to reproduce the data, using the
numerical code of \citet{cerruti2013}, and assuming the EBL
model of \citet{gilmore}. 
The fact that the week-long TeV and X-ray enhanced activity was
detected during a half-year long period of \textcolor{blue}{simultaneous} brightening in the radio, optical, and GeV bands
suggest that all emissions come from the same region, located $\sim$
$10^4$ to $10^5$ Schwarzschild radii away from the black hole, which would point to large-scale emission in PKS 1441+25. This is the first detailed photometric and polarimetric picture of a TeV
FSRQ consistent with a single region being responsible for the multi-band flare. 

The detection of gamma-ray emission from PKS 1441+25 also sets a stringent upper limit on the near-ultraviolet to near-infrared intensity of the EBL, as shown in Figure 5b. The VERITAS and MAGIC  results imply that galaxy surveys have resolved most, if not all, of the sources of the EBL at these wavelengths. Catching a TeV gamma-ray outburst from a quasar provides an excellent opportunity for placing constraints on the diffuse extragalactic background radiation, and offers insights on the relativistic jets of blazars. 

\begin{figure}[t!]
\begin{center}
\includegraphics[scale=0.75]{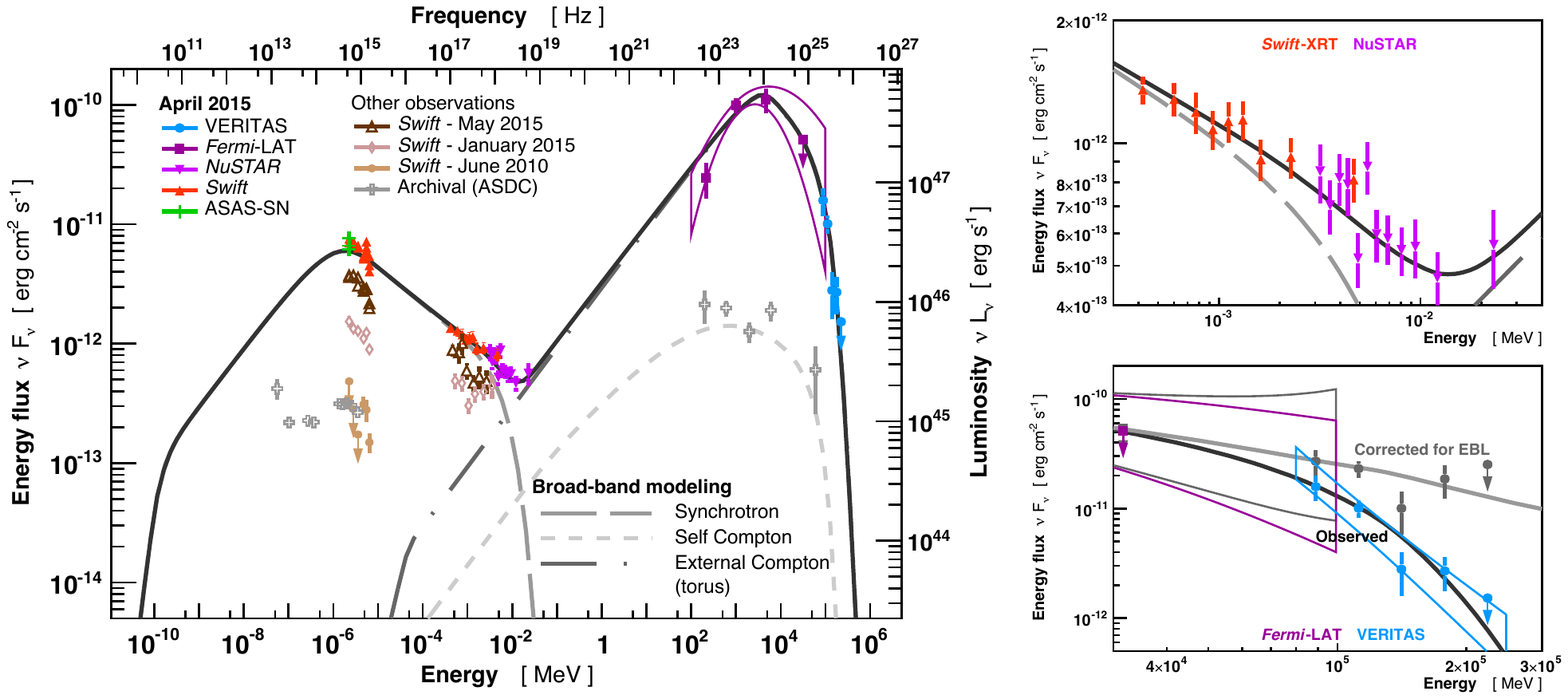} \vskip 0.2in
\includegraphics[scale=0.5]{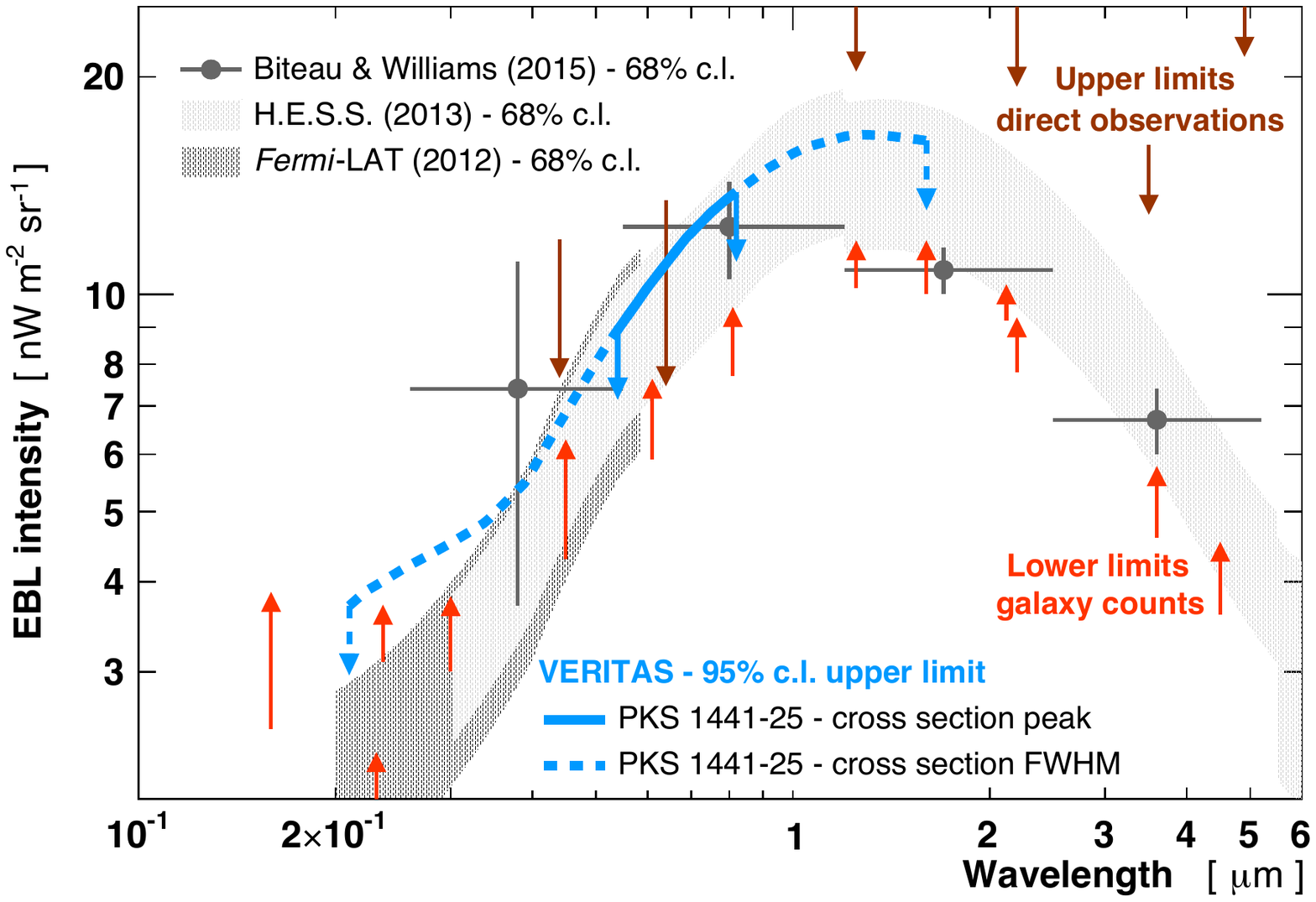} 
\end{center}
\caption{(a) Top: Multiwavelength emission of PKS 1441+25 from X-ray to TeV gamma-ray energies. Broadband model calculations are overlaid~\citep{veritas1441_25}. (b) Bottom: Near-ultraviolet to near-infrared spectrum of the EBL showing the upper limit derived from the VERITAS observations of TeV emission from the distant quasar PKS 1441+25 (shown in blue)~\citep{veritas1441_25}.} 
\end{figure}

\subsection{Extreme Variability in Flux}

Blazars are characterized by extreme and episodic flux variability across all wavelengths, and observations of bright blazar flares result in tighter constraints on the size and location of the gamma-ray emitting region (e.g.~\cite{begelman2008}). 
A bright, short-lived flare in BL Lacertae was measured by VERITAS in June 2011, the first detection of minute-scale variability in a low-frequency peaked BL Lac object.  In this observation, the flux from the source was found to decay by a factor of $\sim 10$ in $13 \pm 4$ min, thus constraining size of emission region to $\sim 2.2\times 10^{13}\delta$ cm, where $\delta$ is the Doppler factor of the jet~\citep{bllac_flare2011}. At the time of the TeV flare, the source was active and variable in GeV gamma rays as detected by the {\sl Fermi}-LAT. Observations with VLBA at 43 GHz also noted the emergence of a radio knot, resolved as moving downstream in the jet, that is likely to be linked to the gamma-ray flare~\citep{bllac_flare2011}. 

Recently, in 2016 October, a fast TeV gamma-ray flare was detected again from BL Lacertae by VERITAS, with a rise time of about 2.3 hours and a decay time of about 36 minutes. Figure 6a shows the VERITAS light curve above 200 GeV for this flare, plotted with  a 4-minute binned light curve. The peak flux from BL Lacerate during this flare was measured to be roughly 180\% the Crab Nebula flux~\citep{bllac_flare2016}. 

 \begin{figure}
\begin{center}
\includegraphics[scale=0.36]{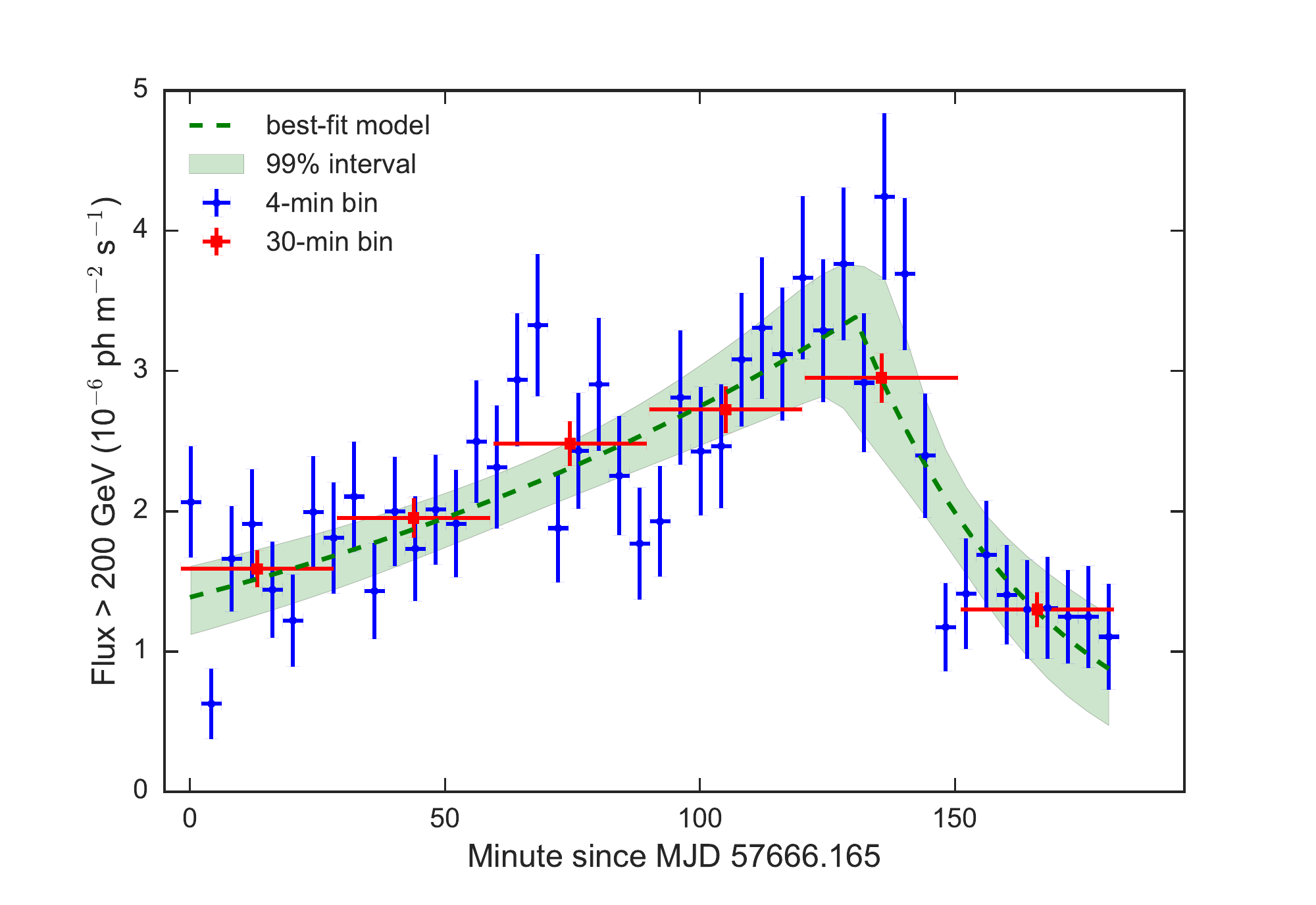} \hfill
\includegraphics[scale=5.5]{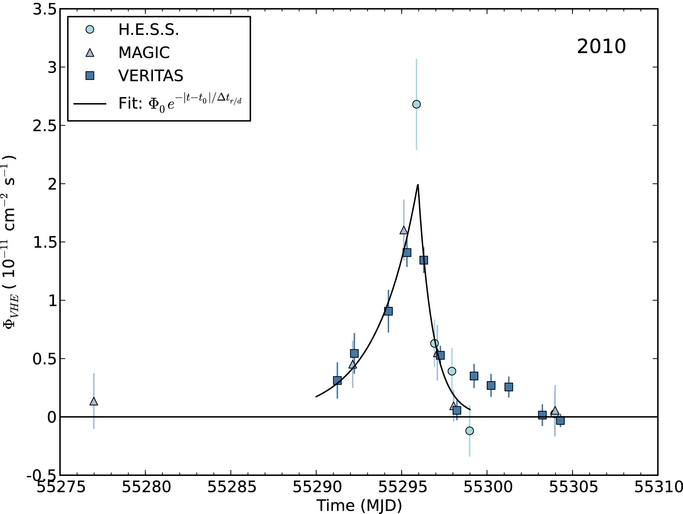}
\end{center}
\caption{(a) Left: VERITAS light curve of BL Lac in October 2016 (energies $> 200$ GeV). Two different time bins are shown, 4-min (blue) and 30-min (red). The green dashed line shows a best fit model using Markov chain Monte Carlo sampling. For details see~\citet{bllac_flare2016}. (b) Right: A bright VHE flare in M 87 in 2010 recorded by VERITAS in a joint campaign with MAGIC and H.E.S.S., showing a variability timescale of $\sim1$ day. The solid curve is the result of the fit of an exponential function to the data~\citep{abramowski2012}.}
\end{figure}

Another AGN that has exhibited several flares at TeV energies is the radio galaxy M 87. Radio galaxies offer an unique opportunity to study VHE emission in close proximity to a super-massive black hole, since they exhibit weak-to-moderate beaming, with jets at larger angles compared to blazar jets. Only two radio galaxies have been detected by VERITAS, M~87 and NGC 1275. Figure 6b shows a bright and isolated VHE flare recorded from M~87 in 2010 by all three IACTs, VERITAS, MAGIC and H.E.S.S.~\citep{abramowski2012}. The flare is characterized by a two-sided exponential function with a rise time of near 1.7 days and fall time of approximately 0.6 days. 

No two blazar flares are similar. VERITAS measured an isolated gamma ray flare in February 2014 from B2 1215+30 that was remarkable because of its extreme luminosity, with the TeV flux reaching 2.4 times the Crab Nebula flux with a variability timescale of less than 3.6 h. The measured flux above 0.2 TeV corresponded to an isotropic luminosity $L_\gamma = 1.7 \times 10^{46}$ erg s$^{-1}$, which was one of the highest to be ever observed from a TeV blazar. The VERITAS observations imply a Doppler factor $\delta> 19$ and place the emitting blob beyond the broad-line region~\citep{veritas_1215}.   The short timescale flux variability observed in B2 1215+30 may be explained as particle acceleration in relativistic shocks or through magnetic reconnection~\citep{sironi2009, giannios2013}. 
These observations suggest a hard spectrum for the electron population with index $p\sim 1.9$, and are consistent with a scenario of magnetic reconnection events in the blazar jet. 

\subsection{Understanding Blazar Spectral Energy Distributions} 

Mulitwavelength (MWL) observations of blazars provide critical input to modeling blazar SEDs and understanding their particle acceleration and  radiation. The key to a successful model calculation is contemporaneous and complete coverage across a broad wavelength range, which has often proved difficult. Dedicated blazar MWL campaigns in recent times have allowed us to make some progress in this area. It is generally found that the SEDs of HBLs are better explained by a one-zone synchrotron self-Compton emission model, although with some degeneracies (for example, see~\cite{boettcher2013} for a review of blazar emission models). Observations of 1ES 0229+200 with VERITAS  over a three year period from 2009 to 2012, provided an SED data set that allowed extensive modeling studies. In particular, with the availability of {\sl Swift}, {\sl Fermi}-LAT, and VERITAS data, it was possible to constrain both the low-energy and high-energy peaks of the SED of the blazar. The SED of the HBL 1ES 0229+200 was found to be compatible with an SSC model, and the data allowed constraining the SSC parameter space, using an algorithm as described by Cerruti et al.~\citep{cerruti2013}. Figure 7 shows the relationship between the Doppler factor $\delta$, with the tangled homogeneous magnetic field $B$, and the energy densities contained in the particles and magnetic field ($U_e$, $U_B$). The latter indicates that the equipartition factor $U_e/U_B$ is between $2 \times 10^4$ and $10^5,$ implying an emission region significantly out of equipartition~\citep{veritas_0229}. 

On the other hand, some blazar SEDs are not well modeled with a simple one zone SSC scenario. For example, observations of the BL Lac object PKS 1440+122 with {\sl Fermi}-LAT, {\sl Swift}, and VERITAS showed that while a synchrotron self-Compton model produces a good representation of the multi-wavelength data, adding an external-Compton or a hadronic component also adequately describes the data~\citep{veritas_1440}. In general, HBL-like SEDs can be modeled by SSC, SSC+EC and lepto-hadronic models. The VERITAS blazar ``long term'' program continues to build up a database of SEDs from a variety of blazars, which will enable focused modeling studies in the future~\citep{wystan_icrc2017}. As we build up an archival data set on VHE blazars, it is possible that high-resolution measurements of blazar spectra at GeV and TeV energies, and variability studies of blazar flares, may help resolve degeneracies between blazar models. 

 \begin{figure}[t!]
\begin{center}
\includegraphics[scale=5.5]{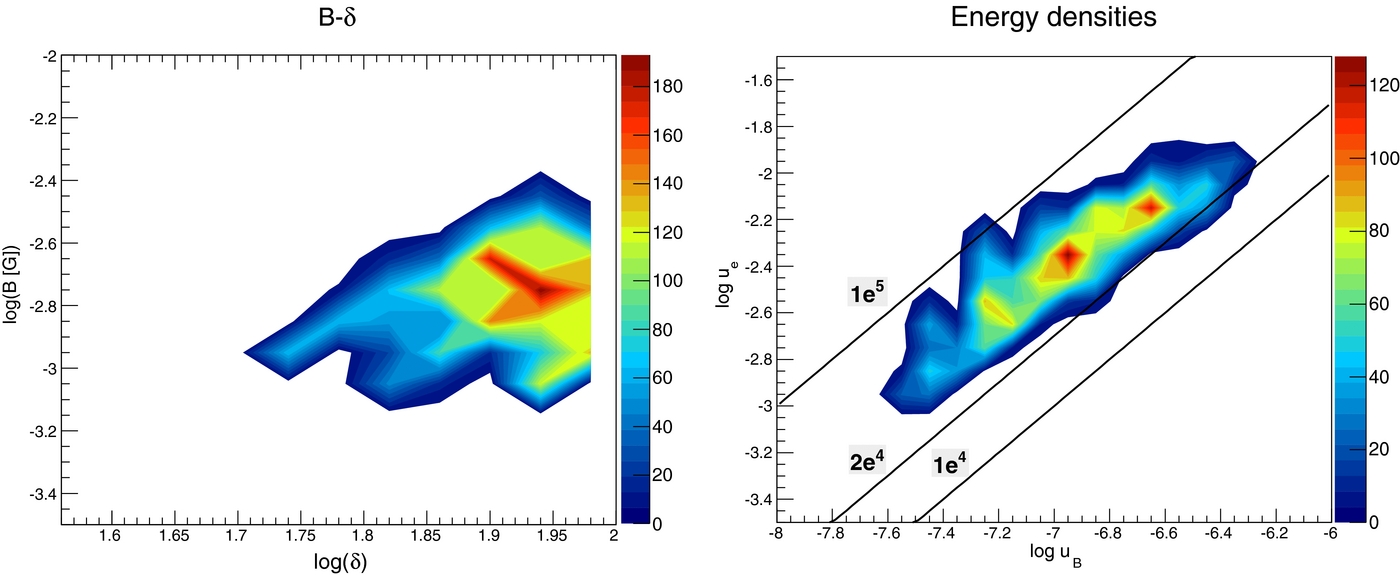} 
\end{center}
\caption{(Left) Constraining the magnetic field $B$ and Doppler factor $\delta$ parameter space obtained from model calculations for the broad band SED of the HBL 1ES 0229+200. (Right) Particle and magnetic energy densities ($U_e$ and $U_B$) parameter space in the model calculation of the SED of 1ES 0229+200. The slanted lines are equipartition contours ($U_e/U_B$). In both plots, the color scale is arbitrary and the most extended contour represents the $1\sigma$ region~\citep{veritas_0229}.} 
\end{figure}

\subsection{New Discoveries: OJ 287}
The TeV sky is rapidly changing, and the very high energy source catalogue continues to grow.  In February 2017, VERITAS reported the detection of the iconic optically bright quasar blazar, OJ 287 \citep{atel_bllac} that shows an interesting 12-year cycle quasi-periodicity in the optical band \citep{shi2007}. OJ 287 is believed to host a binary black hole system at its core \citep{valtonen2006}, and was known previously to be a gamma-ray emitter at GeV energies, with strong flares detected by {\sl Fermi}-LAT.  Strong gamma-ray flares are believed to originate from either the ``core" region or further downstream in the jet \citep{hodgson2017}, but this was the first evidence of VHE emission from the source. Studies are currently underway to understand the TeV emission in the context of multi-wavelength models \citep{bllac_2017}.

\section{ Galactic Astrophysics with VERITAS}

The VERITAS Galactic program encompasses a variety of topics such as the survey of the Cygnus region, study of supernova remnants including non-thermal shells, shell-molecular cloud interactions, and the observation of TeV pulsar wind nebulae (PWNe) associated with high $E_{dot}/d^2$ pulsars, observations of gamma-ray binaries and binary candidates, follow-up of unidentified gamma-ray sources, and focused observations of the Galactic Center. In the following subsections, we highlight a few results from VERITAS. 

\subsection{Supernova Remnants} 

\begin{figure} [b!]
\begin{center}
\includegraphics[scale=0.8]{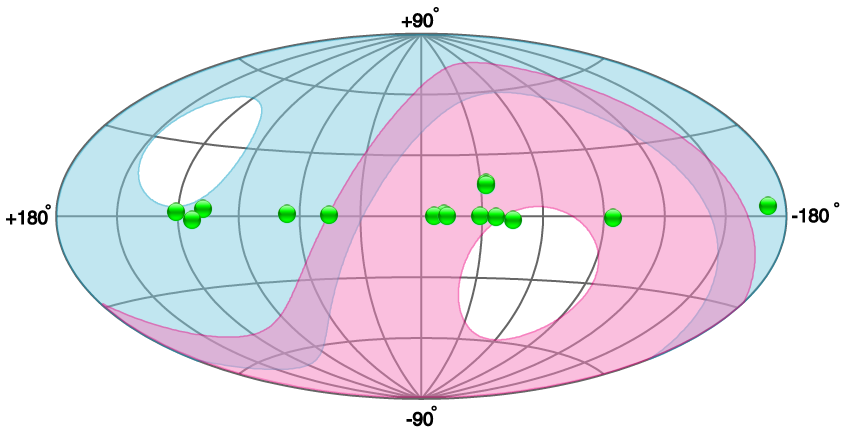} 
\end{center}
\caption{ Skymap showing the SNR shells detected at TeV energies by all three major IACTs,  as of October 2017. The shaded areas show the visibility regions for the northern and southern IACTs; blue for MAGIC/VERITAS and magenta for H.E.S.S. Figure compiled from~\cite{tevcat}.} 
\end{figure}

Supernova remnants are a key component of the VERITAS long term observation plan, since gamma-ray emissions from SNRs offer an opportunity to study cosmic-ray acceleration in Galactic sources. These objects are one of the most violent events in our Universe, with expanding shock waves likely to be capable of accelerating cosmic rays up to multi-TeV energies through diffusive shock acceleration.  Young shell-type SNRs are expected to release a significant fraction of their non thermal output at TeV energies, and several have been detected by current IACTs. Figure 8 shows a skymap of the SNR shells detected at TeV energies in the last several years, by H.E.S.S., MAGIC and VERITAS~\citep{tevcat}. All the SNRs detected at TeV energies are remnants of recent supernova explosions, less than a few 1000 years old, except for IC 443 which is a middle-aged remnant. Several of the TeV SNRs have been resolved in VHE gamma rays, and H.E.S.S. has detected at least five SNRs with shell type morphology~\citep{hess_snr_review}, RX J0852.0$-$4622 (Vela Junior), SN 1006, HESS J1731-347, RX J1713.7-3946, and RCW 86. In each case, there is clear correlation between non-thermal X-ray and VHE gamma-ray emissions, giving us the chance to investigate sites of CR acceleration. {\sl Fermi}-LAT data show that SNRs are firmly established as sources of cosmic rays, with the spectrum peaking in the GeV range, and consistent with a pion-bump. Figure 9 shows the gamma-ray spectrum of IC 443 measured with {\sl Fermi}-LAT, with TeV data from MAGIC, and VERITAS overlaid~\citep{fermi_ic443}. The {\sl Fermi}-LAT SED shows the characteristic cutoff around 200 MeV, which is a direct evidence of hadronic interactions, with the gamma-ray spectrum tracing the source spectrum of cosmic rays. 

 \begin{figure}[t!]
\begin{center}
\includegraphics[scale=0.40]{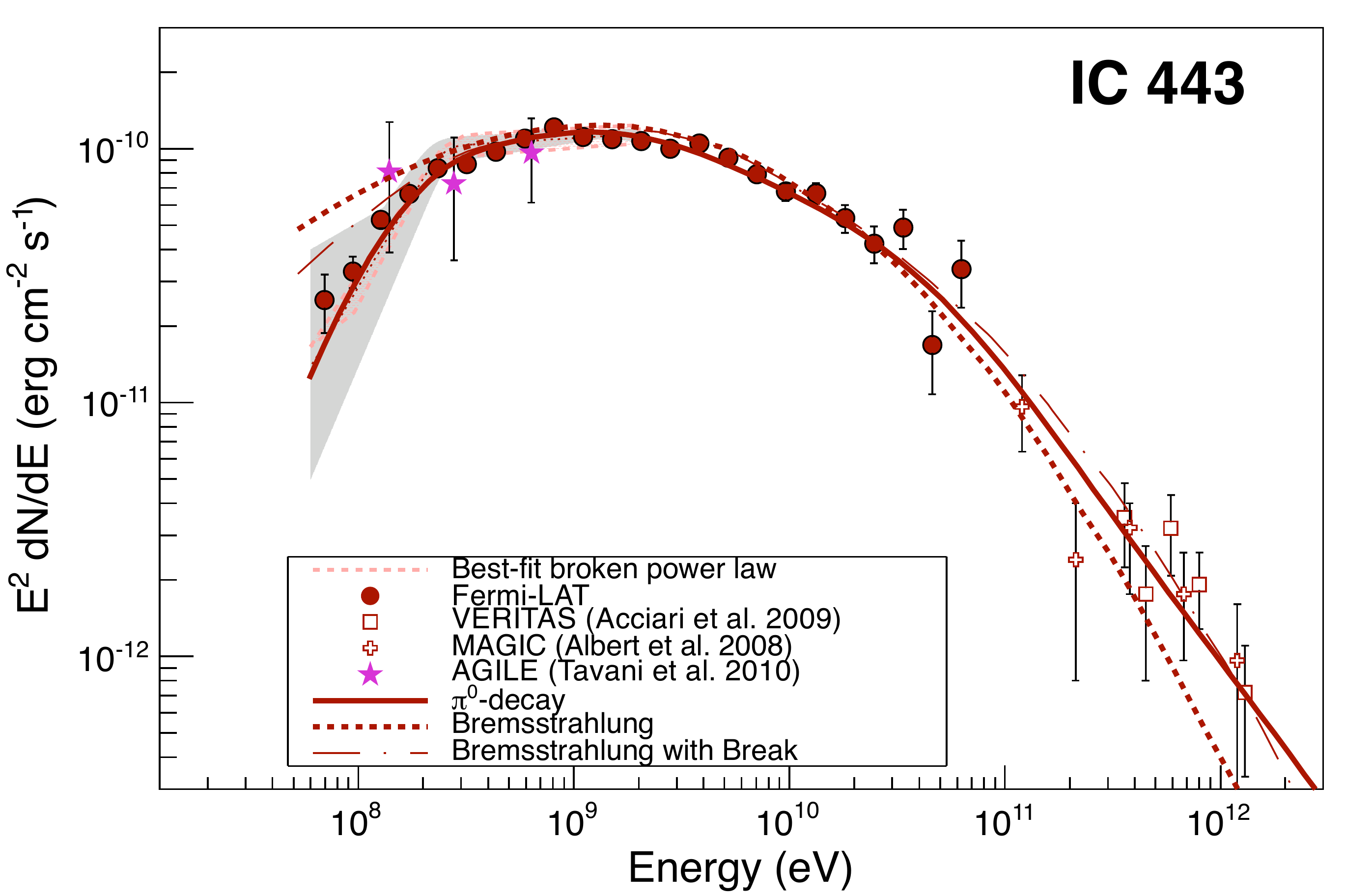} 
\end{center}
\caption{Gamma-ray spectrum of IC 443 as measured with {\sl Fermi}-LAT, showing the characteristic feature at $\sim 200$ MeV, which is referred to as the ``pion-decay bump.'' The {\sl Fermi}-LAT data are well fit with a model calculation, where the gamma rays come from the decay of neutral pions generated in proton-proton or nuclear-nuclear collisions. Also shown overlaid are data from MAGIC and VERITAS at TeV energies. See \citet{fermi_ic443} for details on all modeling parameters.} 
\end{figure}

Cas A and Tycho are two SNRs detected by VERITAS. Both sources are unresolved but comparatively bright at TeV energies. With long exposures, VERITAS has been able to make precise spectral measurements for both sources~\citep{holder_heidelberg2016}. The Tycho SNR provides a particularly good target for investigating hadronic cosmic-ray interactions as it is located in a relatively clean environment, and because it is a young type Ia SNR that is well studied in other wavelengths. The Canadian Galactic Plane Survey (CGPS) shows evidence of a molecular cloud in the north east corner of the remnant, providing a target for possible hadronic interactions~\citep{cgps_tycho}.  At X-ray energies of 4.0 to 6.0 keV, Chandra data show non-thermal emission features and shock fronts, with evidence for electrons with energies up to 100 TeV. 

\begin{figure}[h!]
\begin{center}
\includegraphics[scale=1.25]{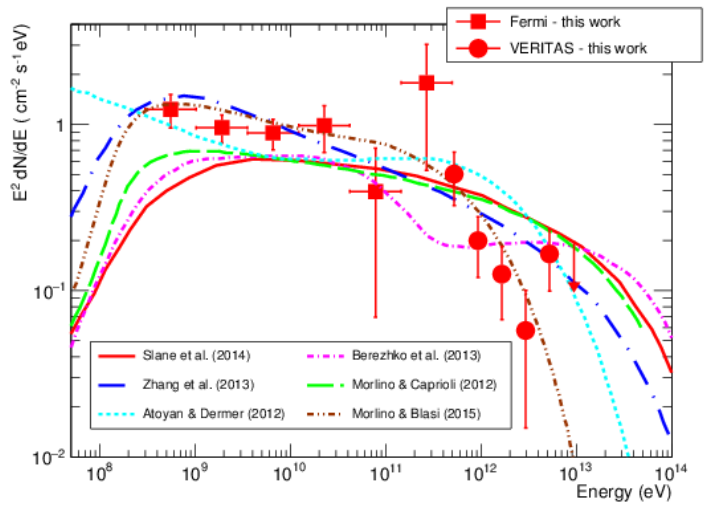} 
\end{center}
\caption{ Spectral energy distribution of Tycho showing data from VERITAS and {\sl Fermi}-LAT, with overlaid models, as described in~\citet{veritas_tycho2017}.} 
\end{figure}

VERITAS recently reported results from 147 hours of observations on Tycho's SNR, spanning  data from 2008 to 2012~\citep{veritas_tycho2017}, which is detected as a point source at VERITAS energies. Enhanced sensitivity of the upgraded VERITAS cameras allowed the measurements to extend down to near 400 GeV. The spectrum of Tycho measured by VERITAS  is consistent with a power-law with a gamma-ray photon index $\Gamma = 2.92\pm0.42_{\rm stat}\pm0.20_{\rm sys}$. A broad band study of the SED of Tycho was carried out, including data at MeV and GeV energies from {\sl Fermi}-LAT. Figure 10 shows the SED of Tycho at VERITAS and {\sl Fermi}-LAT energies.  The VERITAS spectrum is found to be softer than that measured at {\sl Fermi}-LAT energies, and the current VERITAS results indicate that the maximum particle energy in Tycho may be lower than previously suggested. Since the first TeV detection of Tycho~\citep{veritas_tycho2011}, several emission models have been proposed to explain its high energy behavior (see for example~\citet{slane_tycho}). Fig. 10 shows some of these model calculations overlaid on the new VERITAS data, and are found to be inadequate in explaining the new VERITAS data. One of  the models in Fig. 10 ``Morlino \& Blasi 2015'' presents a good fit to the data, assuming the presence of neutral hydrogen close to the shock and that the gamma-ray emission is largely in the dense northeastern region of the remnant~\citep{blasi2016}. 

IC443 is an older SNR and is one of the classic examples of a supernova remnant interacting with a molecular cloud in an inhomogeneous environment. VERITAS has recently been able to resolve the gamma-ray emission from the entire shell of the remnant, as shown in Figure 11~\citep{ic443_veritas}, after a deep observation campaign. The spatial morphology of the VERITAS emission correlates well with the GeV emission from {\sl Fermi}-LAT~\citep{fermi_ic443}, and shows gamma-ray emission at the position of the brightest maser as well as from the entire northeast lobe. Along with the SNR W51C, detected by MAGIC~\citep{w51}, in the W51 complex, IC 443 is one of two extended northern SNR to be resolved at gamma-ray energies. 
IC 443 offers the opportunity to study a system where the expanding shock wave is interacting with gas and clouds. The spatial resolution of IC 443 by VERITAS has been one of the highlight results from VERITAS. Other resolved TeV SNRs are in the southern hemisphere, and have been detected at TeV energies by H.E.S.S.~\citep{hess_1713}. 

 \begin{figure}[t!]
\begin{center}
\includegraphics[scale=0.3]{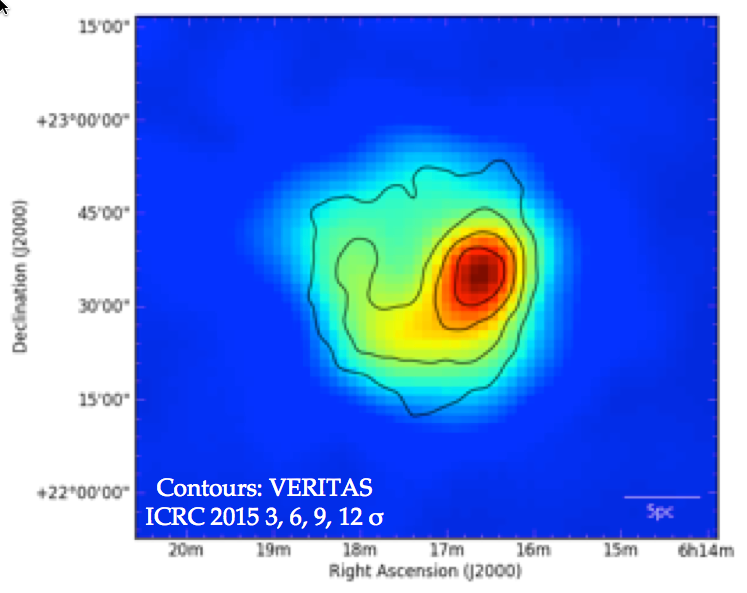} 
\end{center}
\caption{ {\sl Fermi}-LAT counts map of gamma-ray photons from IC 443 selected above 5 GeV, overlaid with VERITAS significance contours for the detection of IC 443. The {\sl Fermi}-LAT map is derived from 83 months of data. VERITAS detects gamma-ray emission over the full extent of the supernova remnant~\citep{ic443_veritas}.} 
\end{figure}

\subsection{VERITAS Studies of Gamma-Ray Binaries} 

Binaries are the only variable and point-like sources in our galaxy, and whether the particle acceleration in these systems happens in jets or colliding winds is still an open question. TeV emission probes the highest energy particles accelerated. VERITAS recently reported on the long-term TeV observations of the gamma-ray binary HESS J0632+057~\citep{veritas_0632}, a source that has now been observed nearly ten years by H.E.S.S., MAGIC and VERITAS. The source was found to have a period of $315\pm 5$ days, derived from X-ray data~\citep{swift_0632}, and VERITAS has observed the source through nearly its entire orbital period. VERITAS and {\sl Swift} observations show that the gamma-ray and X-ray fluxes are correlated (Fig. 12a), which can be explained by a simple one-zone leptonic emission model, where relativistic electrons lose energy by synchrotron and inverse Compton emission~\citep{veritas_0632}. 

 \begin{figure}[t!]
\begin{center}
\includegraphics[scale=0.45]{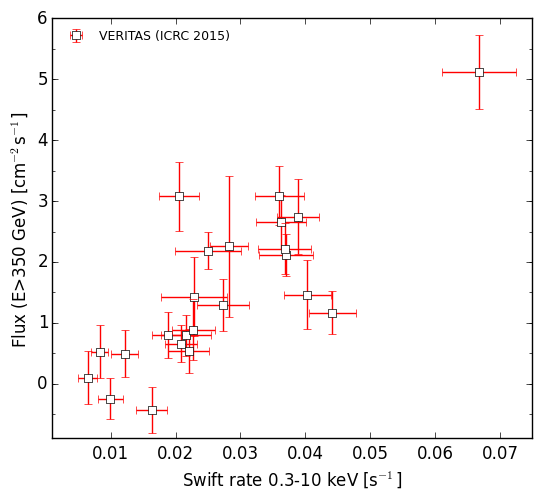} \hfill
\includegraphics[scale=0.23]{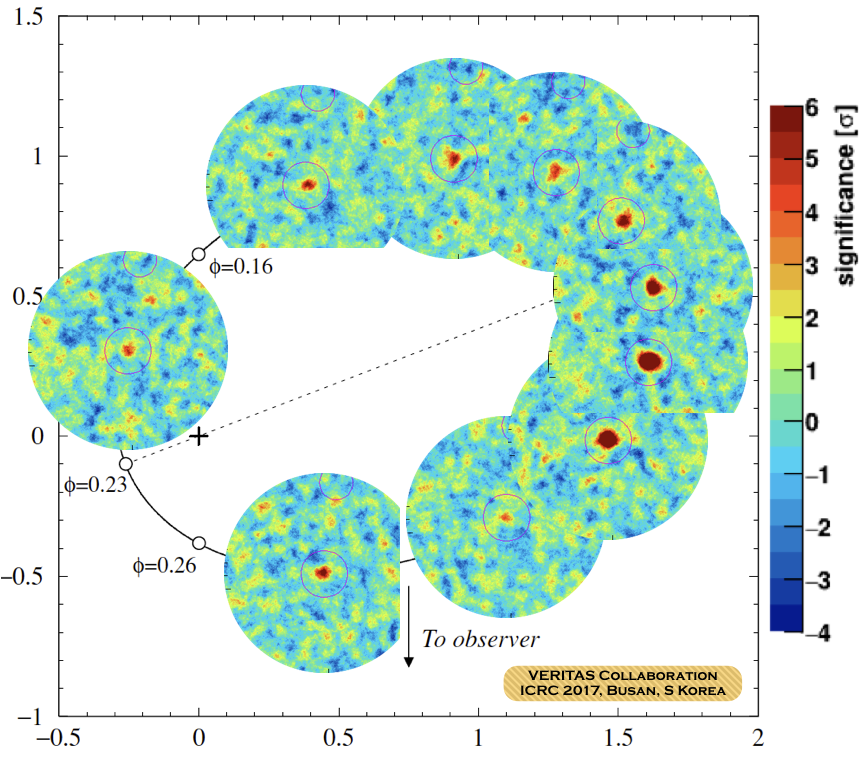} 
\end{center}
\caption{(a) Left: VERITAS gamma-ray ($>350$ GeV) fluxes as a function of {\sl Swift} X-ray (0.3-10 keV) count rates for contemporaneous observations (defined here as $\pm 2.5$ day intervals around the date of the gamma-ray observations) for HESS J0632+057~\citep{veritas_0632}. (b)Right: A compilation of VERITAS observations of LS I$+61^\circ$303 for different orbital phases, with the time progression being shown clockwise. The source is detected by VERITAS at $> 5\sigma$ significance at all phases except in the phase bin 0.4 to 0.5. Figure from \citet{kar_lsi}.} 
\end{figure}

Another well-studied binary at gamma-ray energies is the high-mass X-ray binary (HMXB) LS I+61$^\circ$303, which has been detected across all wavelengths from radio to TeV, and has been found to exhibit a high degree of flux modulation over a single orbit of period approximately 26.5 days~\citep{kar_lsi}. The system consists of a massive B0 Ve star and a compact object, but the nature of the central object, whether it is a black hole or a neutron star has not yet been resolved. A search for pulsations at radio, X-ray, or GeV bands has not found any evidence of pulsed emission. LS I +61$^\circ$ 303 has been extensively observed by VERITAS. 
Figure 12b shows skymaps of LS I +61$^\circ$ 303 obtained from ten years of VERITAS observations of the source, for ten different phase bins, showing maximum flux in the 0.55 to 0.65 phase range, during its apastron pasage. This data set of more than 200 hours will be useful in determining the nature of the unknown compact object (neutron star or microquasar) in the source. It has been suggested that the baseline TeV emission and VHE outbursts near apastron could be explained by the so called neutron star flip flop model \citep{torres2012}. It is interesting to note that LS I+61$^\circ$ 303 also exhibits a super-orbital period of about 4.5 years  in radio, X-ray and GeV emission, and  TeV gamma rays \citep{superorbital}. The cause of the super-orbital periodicity is not completely known, but the detection of TeV emission is consistent with the predicted long-term behavior of the flip-flop model~ \citep{superorbital}. 

The VHE binary population remains small, with only 6 binaries known to emit at TeV energies~\citep{tevcat}. TeV J2032+4130 may turn out to be a new TeV binary, associated with the pulsar/Be-star binary system PSR J2032+4127/MT91 213, with the pulsar assumed to be in a long period (45-50 years), highly eccentric orbit with MT91 213~\citep{lyne}. With the predicted periastron in November 2017, VERITAS and MAGIC observations were carried out in Fall 2017, and both collaborations recently announced the detection of a point source VER J2032+414 at the location of PSR J2032+4127~\citep{atel2032}. This observing campaign is just beginning, and it will be interesting to see if targeted observations continue to find enhanced VHE emission.

\subsection{The Galactic Center} 

The Galactic Center is at large zenith angles for VERITAS, and so VERITAS is able explore the region at energies greater than 2 TeV. This is a rich region for gamma-ray studies, revealing diffuse emission, as well as the composite supernova remnant G0.9+0.1, and the supermassive black hole Sgr A*. Recent studies by H.E.S.S.  show diffuse gamma-ray emission in the Galactic Center region, suggesting a source accelerating cosmic particles to PeV ($E > 10^{15}$ eV) energies~\citep{hess_galactic}. Figure 13a shows the skymap from VERITAS, following a deep observing program of the region between 2010 and 2014. Diffuse emission along the Galactic ridge is evident, after the point sources SgrA* and SNR G0.9+0.1 are subtracted out~\citep{veritas_gc}. 
The VERITAS data, as shown in Fig. 13b, offer improved statistics at multi-TeV energies (as a result of larger effective areas for large zenith angle observations), and together with the H.E.S.S. measurements with excellent statistics at lower energies, provide a rich data set for joint spectral studies between 0.2 to 50 TeV. VERITAS also reported on the detection of a new source VER J1746-289 at $>2$ TeV along the Galactic plane, which looks point-like (Fig. 13a, top panel), and could be related to a combination of local enhancement in the diffuse emission near the Galactic Center and a known H.E.S.S. source in the region \citep{hess_galactic}.  
One of the outstanding questions about this region is a definitive explanation of the gamma-ray emission from Sgr A*. Further studies of this region will help resolve some of these outstanding questions. 

\subsection{The Cygnus Region}  

 \begin{figure}[b!]
\begin{center}
\includegraphics[scale=0.18]{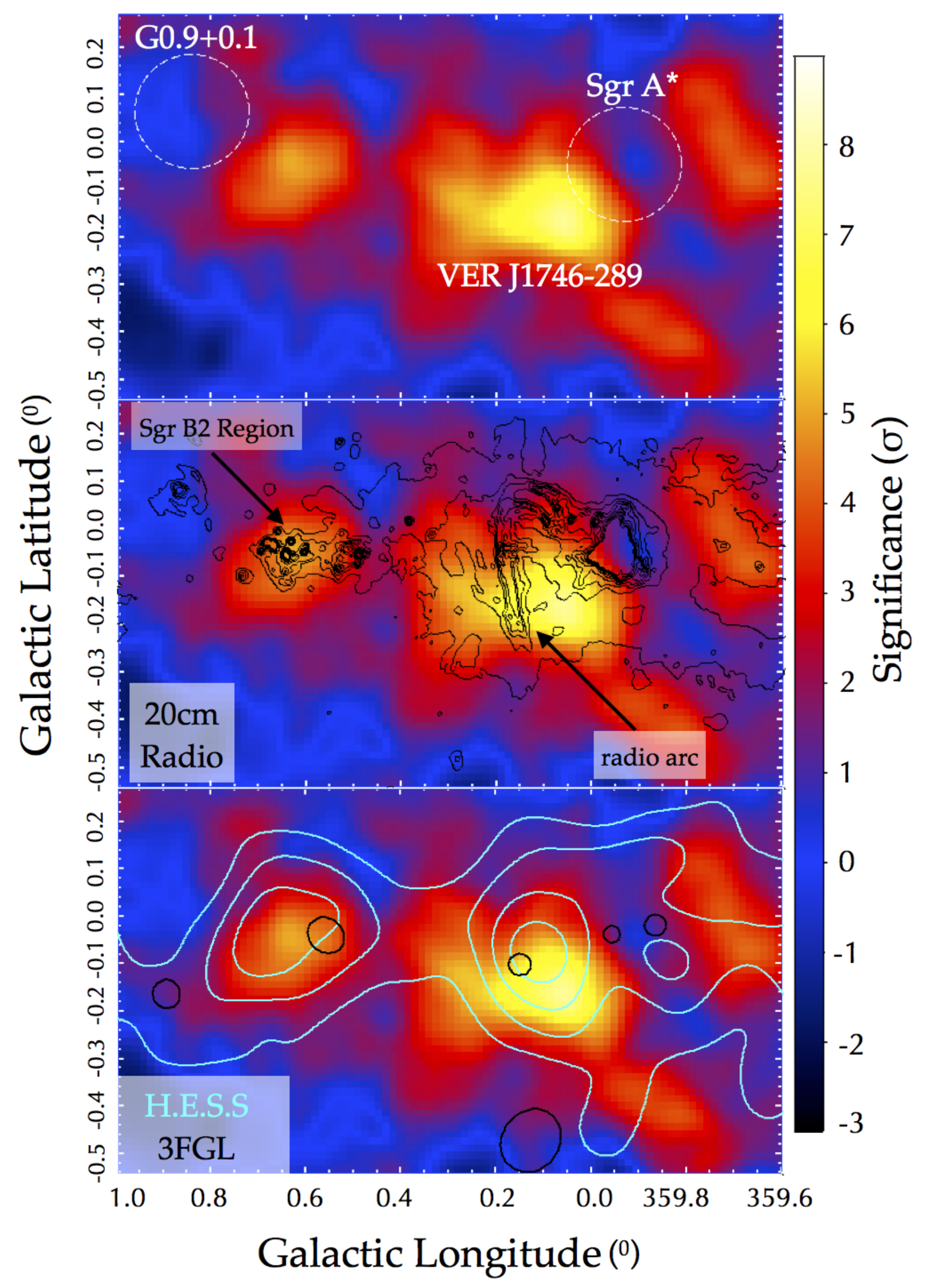}\hfill
\includegraphics[scale=0.12]{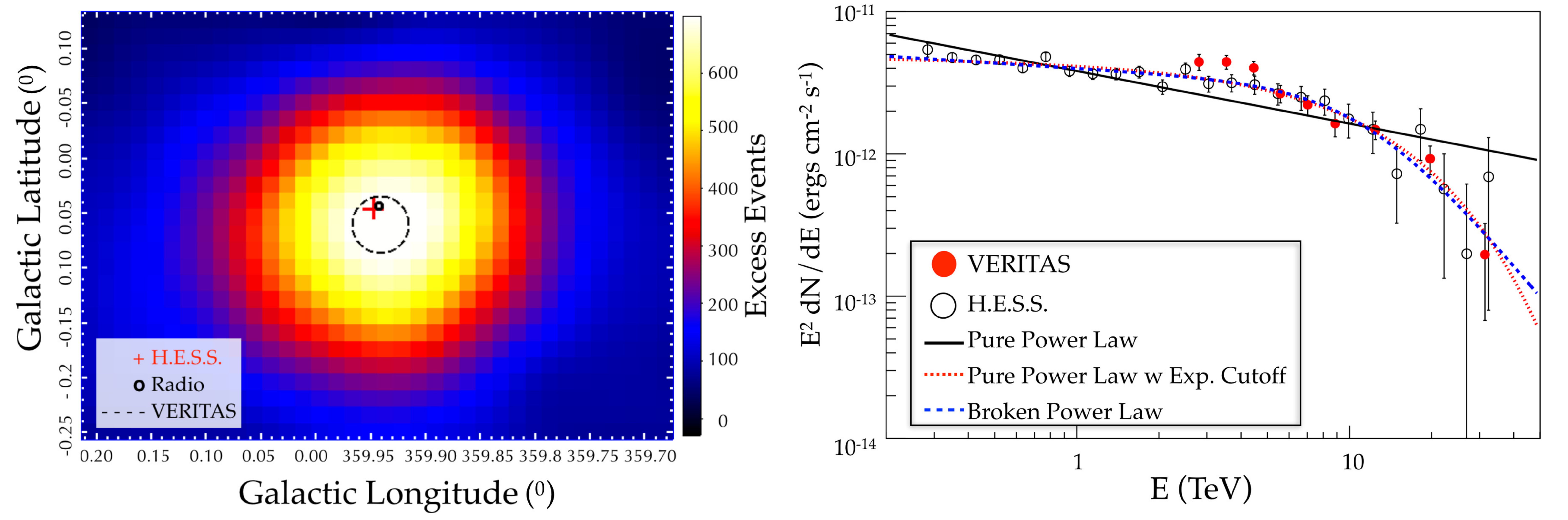} \vfill
\end{center}
\caption{(a) Top: The VERITAS $>2$ TeV significance maps of the Galactic Center region after subtracting excess emission from Sgr A* and G $0.9+0.1$. The top panel shows the locations of the subtracted point sources as well as the VERITAS source VER~J1746-289. VLA 20cm radio contours are shown in the middle panel, with H.E.S.S. excess event contours and {\sl Fermi}-LAT 3FGL sources shown in the bottom panel. (b) Bottom: The differential energy spectrum of Sgr A*, as measured by VERITAS and H.E.S.S.. The lines describe various power-law model fits. Figures and captions are from~\cite{veritas_gc}.} 
\end{figure}

VERITAS carried out an observation of the Cygnus region from 2007 through 2012, accumulating more than 300 hours of data. The Cygnus region  is a natural laboratory for the study of cosmic rays and their origins, as it is the largest and most active region of creation and destruction of massive stars in the Milky Way. A comprehensive report on the Cygnus region by VERITAS is under preparation, and preliminary results have been presented at the 35th ICRC~\citep{cygnus_veritas}. Figure 14 shows a skymap of TeV emission measured by VERITAS of this complex region, showing four VERITAS sources VER J2019+407, VER J2031+415, VER J2016+371, and VER J2019+368~\citep{bird_icrc2017}. The VERITAS survey of the Cygnus region has roughly similar sensitivity as the H.E.S.S. Galactic plane survey (GPS)~\citep{hess_gps}. An analysis of seven years of {\sl Fermi}-LAT data was also carried out to compare with the VERITAS results. A multi-wavelength interpretation of these results is currently under way~\citep{bird_inprep}.

 \begin{figure}[b!]
\begin{center}
\includegraphics[scale=0.7]{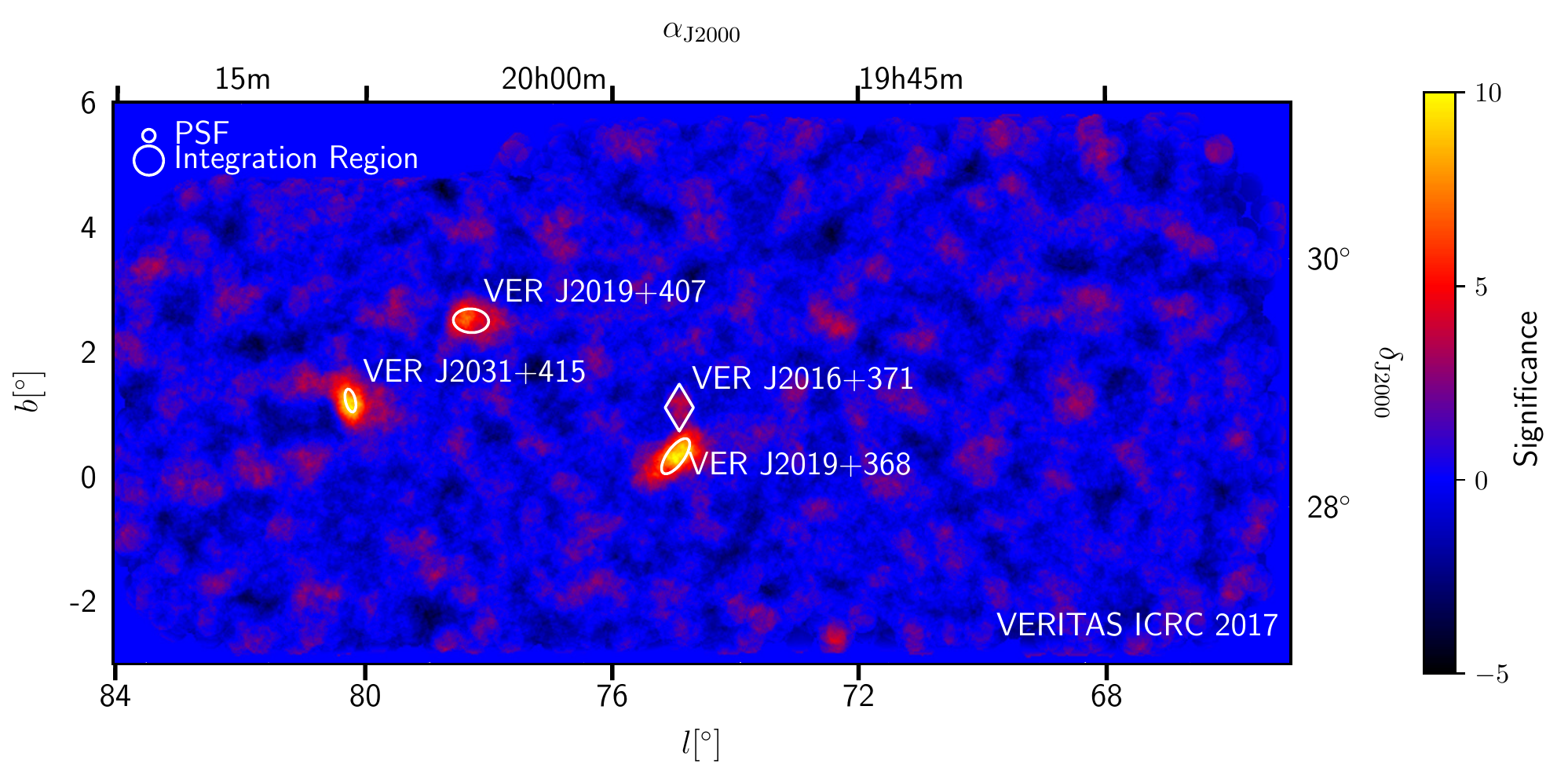}
\end{center}
\caption{VERITAS VHE gamma-ray significance map of the Cygnus region at energies $> 100$ GeV~\citep{bird_icrc2017}. The survey region is a $15^\circ$ by $5^\circ$ portion of the Cygnus region centered on ($l=74.5^\circ$, $b=1.5^\circ$).}
\end{figure}

\subsection{The Crab Pulsar}

The MAGIC Collaboration reported the first detection of pulsed gamma-ray emission above 25 GeV from the Crab pulsar (and the first detection of a pulsar $>25$ GeV), suggesting that the emission zone is far out in the pulsar magnetosphere \citep{magic_crab}. 
The first detection of pulsed emission above 100 GeV from the Crab pulsar was reported by VERITAS, a result  that was difficult to explain on the basis of present pulsar models \citep{OtteScience2011}. The VERITAS data indicate that the very high energy gamma-ray emission above 100 GeV is unlikely to be primarily due to curvature radiation. These results imply that the gamma rays are not produced at the inner acceleration gap, but rather close to or outside the light cylinder, beyond 10 stellar radii from the neutron star. Furthermore, MAGIC reported the first detection of pulsed emission from the Crab pulsar above 400 GeV with the spectrum reaching up to 1.5 TeV \citep{magic_2tev}, challenging emission models. 
An updated VERITAS analysis on the Crab pulsar was presented by \citet{nguyencrab} using 194 hours of quality-selected data from VERITAS, searching for pulsed emission beyond 400 GeV, but there was no significant detection. 
Studies of terraelectronvolt pulsed emission from pulsars may provide the opportunity to probe physics at the Planck scale and constrain LIV (Lorentz Invariance Violation) effects (see for example \citet{OtteICRC32}. A study constraining LIV using the TeV Crab pulsar emission was recently carried out by MAGIC \citep{magic_liv}.

\section { Fundamental Physics Studies and Cosmology} 

VERITAS data has contributed to our understanding in several areas of fundamental physics and cosmology. We highlight a few particular areas of study in the following subsections. 

\subsection{Dark Matter} 

VERITAS has carried out a program for indirect dark matter searches to look for weakly interacting massive particles (WIMPs) in the mass range of $\sim$ 100 GeV to 10 TeV. WIMPs are well motivated dark matter candidates in extensions of the Standard Model of particle physics (supersymmetry, Kaluza-Klein). The neutralino, the lightest SUSY particle, could self-annihilate to produce gamma rays. Dwarf spheroidal galaxies (dSphs) are attractive targets for indirect dark matter searches as they are nearby (20 to 200 kpc) and have large mass to light ratios. Additionally, these sources are not known to be high energy gamma-ray sources, and therefore are not expected to produce a confusing astrophysical signal.
IACTs such as H.E.S.S., MAGIC or VERITAS are able to put stringent constraints at the high mass range ($\geq 1$ TeV), where instruments such as {\sl Fermi}-LAT do not have significant sensitivity. 
 VERITAS recently published results on 230 hours of data taken on five dwarf galaxies, Bo\"otes I, Draco, Segue I, Ursa Minor, Willman I, observed between 2007 and 2013~\citep{dwarf_veritas2017}, presenting constraints on the annihilation cross section of WIMP dark matter. VERITAS data show no evidence of gamma-ray emission from any individual dwarf spheroidal galaxy, nor from a joint analysis of the four dwarfs Bo\"otes I, Draco, Segue I and Ursa Minor. Willman I was not included in the joint analysis, since some studies have shown evidence of irregular kinematics in the stellar population of Willman I, and its $J$ factor cannot be reliably calculated~\citep{wilman2011}. The paper quotes the following result: The derived upper limit on the dark matter annihilation cross section from the joint analysis is $1.35\times10^{-23}$ cm$^{3}$s$^{-1}$ at 1 TeV for the bottom quark $(b\bar b)$ final state, $2.85\times10^{-24}$ cm$^{3}$ s$^{-1}$ at 1 TeV for the tau lepton $(\tau^+\tau^-)$ final state, and $1.32\times10^{-25}$ cm$^3$ s$^{-1}$ at 1 TeV for two photon final state~\citep{dwarf_veritas2017}. 
MAGIC has also recently published results on annihilation cross section limits using 160 hours of data on one dwarf spheroidal galaxy Segue 1, for the $b\bar b$ and $\tau^+ \tau^-$ channels, which are the most constraining limits to date~\citep{magic_DM}. 
 Figure 15 shows the expected median velocity-weighted annihilation cross section limits from the joint analysis of all five dwarf galaxies for all different possible channels. As shown in the figure, the strongest continuum constraints are from a heavy lepton final state. The current upper limits on the annihilation cross section are still about two orders of magnitude away from the value of the relic abundance. One hopes that future improvements in instrumental sensitivity might bridge the gap.

\begin{figure}[t!]
\begin{center}
\includegraphics[scale=1.0]{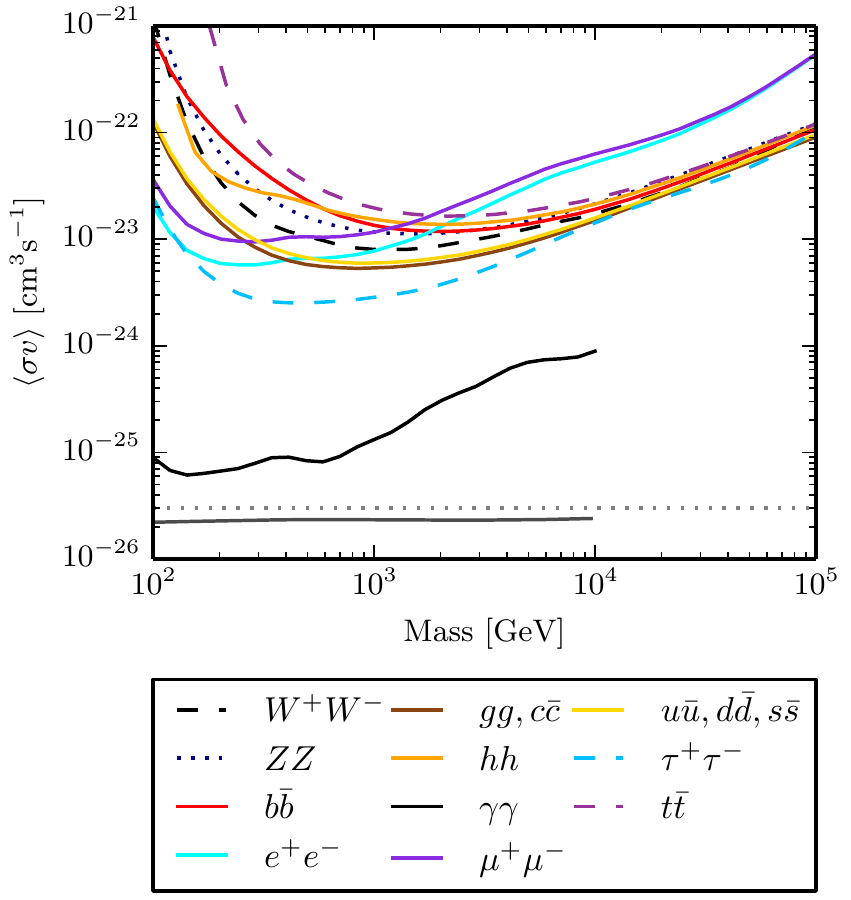} 
\end{center}
\caption{Upper limit calculations  from VERITAS observations of dSphs on the weighted annihilation cross-section $<\sigma v>$ as a function of WIMP mass. The median annihilation cross section limit from all dwarf galaxies and for all channels are shown. The thin dashed horizontal line corresponds to the benchmark value of the required relic abundance cross section $(3 \times 10^{26}$ cm$^3$ s$^{-1}$), while the solid horizontal line corresponds to the detailed calculation of the same quantity~\citep{dasgupta}. Figure from \cite{dwarf_veritas2017}.}
\end{figure}

\subsection{Constraints on the intergalactic magnetic fields} 

VERITAS has carried out a search for magnetically-broadened emission from blazars, which could potentially be detected if there is a non-zero magnetic field in the inter galactic medium. 
Such an intergalactic magnetic field (IGMF) could lead to cascade emission in blazars, particularly for the extreme-HBLs with hard spectrum. 
Characterizing the IGMF could help understand the origin and evolution of the primordial magnetic fields. A recent study presented the latest VERITAS results on the search for extended gamma-ray emission. Based on observations on a number of strongly-detected TeV blazars at a range of redshifts~\citep{elisa_igmf}, the research found no indication of angularly broadened emission. In particular, for the hard-spectrum blazar 1ES 1218+304, an IGMF strength of $5.5\times10^{-15}$ G to $7.4 \times 10^{-14}$ G was excluded at the 95\% CL, assuming an EBL model of \citet{gilmore}, spectral index $\Gamma =1.66$ and cutoff energy $E_C=10$ TeV. Results have also been published by the H.E.S.S. Collaboration, excluding IGMF strengths of $(0.3-3)\times 10^{-15}$ G at the 99\% confidence level, using data from three blazars 1ES 1101-232, 1ES 0229+200 and PKS 2155-304~\citep{hess_igmf}.

\subsection{Electron Spectrum}

At TeV energies, cosmic-ray electrons provide a direct measurement of local cosmic-ray acceleration and diffusion in our Galactic neighborhood since they lose energy quite rapidly via inverse Compton scattering and synchrotron processes. VERITAS, similar to the other IACTs, is able to measure the combined spectrum of the cosmic-ray electrons and positrons beyond the energy range explored by {\sl Fermi}-LAT and AMS, although it is not able to distinguish between electrons and positrons. Figure 16 shows the preliminary VERITAS cosmic-ray electron energy spectrum in the energy range from 300 GeV to about 5 TeV~\citep{crelectron_veritas2017}. The spectrum is best fit with a power-law of two indices, $-3.1\pm 0.1_{\rm stat}$ below, $-4.1\pm 0.1_{\rm stat}$ above, and a break energy of $710\pm 40$ GeV. Note that the uptick in the final VERITAS data point is within $2\sigma$ of the best fit line, and should therefore not be over-interpreted. The H.E.S.S. data are also within their systematic limits (not shown here), and are consistent with the VERITAS results. Recently H.E.S.S. has reported the detection of electron-like events extending to about 20 TeV, the highest energies detected to date. These results, reported at the 2017 ICRC, show a spectrum that can be fitted with a smooth, broken power law, with no features seen up to the highest energies \citep{hess_electron_icrc2017}. 

\begin{figure}[t!]
\begin{center}
\includegraphics[scale=0.7]{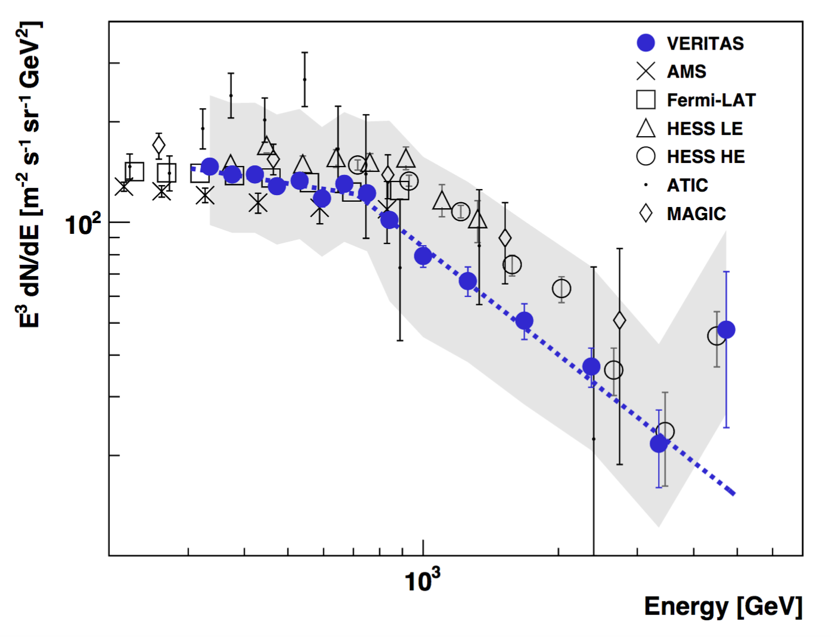} 
\end{center}
\caption{A preliminary cosmic-ray electron spectrum as a function of energy measured by VERITAS. The VERITAS data are shown as solid blue circles. For comparison, archival data from other experiments in the same energy range are included. The systematic uncertainty of the VERITAS measurements is shown as a gray band. See~\cite{crelectron_veritas2017} for references. }
\end{figure}

\section{Multimessenger \& New Partners} 

As of this writing, VERITAS has started collaborative work with both the HAWC and IceCube Collaborations, and has actively followed up on LIGO/VIRGO alerts. These joint studies provide new avenues using multi-messenger techniques to explore the highest energy events in the Universe. 

VERITAS has carried out follow up studies of TeV gamma-ray sources from the second HAWC catalog~\citep{hawc_cat}, that lists thirty nine very high energy gamma-ray sources based on 507 days of exposure time. VERITAS and {\sl Fermi}-LAT observations of these fields have seen recently been reported by \citet{park_hawc}. VERITAS has also carried out a search for TeV gamma-ray emission associated with IceCube high-energy neutrinos, in the hope of finding hadronic cosmic-ray accelerators, which would be sources of both gamma rays and neutrinos.  Over the last two years, VERITAS has carried out observations of several muon neutrino events, since they have better angular reconstruction uncertainty ($<1^\circ$) than other neutrino events. Figure 17a shows the sky map of high-energy neutrino events as measured by IceCube~\citep{icecube_skymap}. VERITAS can view the northern sky (shown in blue). The positions of the IceCube contained muon tracks are highlighted with  red circles (C5, C13, and C37). VERITAS carried out observations of the three muon track events, but did not see a gamma-ray excess. Figure 17b shows the skymap of VERITAS observations of C5 (see~\citet{icecube_veritas} for details). VERITAS has also carried out target-of-opportunity observations on the blazar TXS 0506+056, seen flaring in {\sl Fermi}-LAT~\citep{fermi_icecube}, within the localization region of the IceCube neutrino IC 170922A~\citep{ic170922}. It is interesting to note that MAGIC reported on detecting this blazar during the time window of the {\sl Fermi}-LAT observations~\citep{magic_icecube}. 

Responding to transients alerts is of the highest priority for VERITAS, and the goal is to search for sources that emit in two or more ``cosmic messenger'' channels (photons, neutrinos, cosmic rays, and gravitational waves). VERITAS has carried out prompt follow-up observations of LIGO/VIRGO alerts, when possible. Recently, VERITAS reported follow up observations of the localization region for the gravitational-wave candidate G268556 in January 2017~\citep{veritas_gcn}. Thirty nine consecutive exposures were taken to tile the localization region of the northern section of the 50\% containment region for the event, which was observable at  $>50^\circ$ elevation from the VERITAS site. The exciting  LIGO/VIRGO event GW170817 in August 2017~\citep{ligo170817}, for which there were detections of electromagnetic counterparts, was unfortunately not unobservable for VERITAS due to its location in the southern hemisphere. (VERITAS is also shut down in August due to the monsoon season in Arizona). In the future VERITAS will make it a priority to follow up such events in its field of view.

 \begin{figure}[b!]
\begin{center}
\includegraphics[scale=0.26]{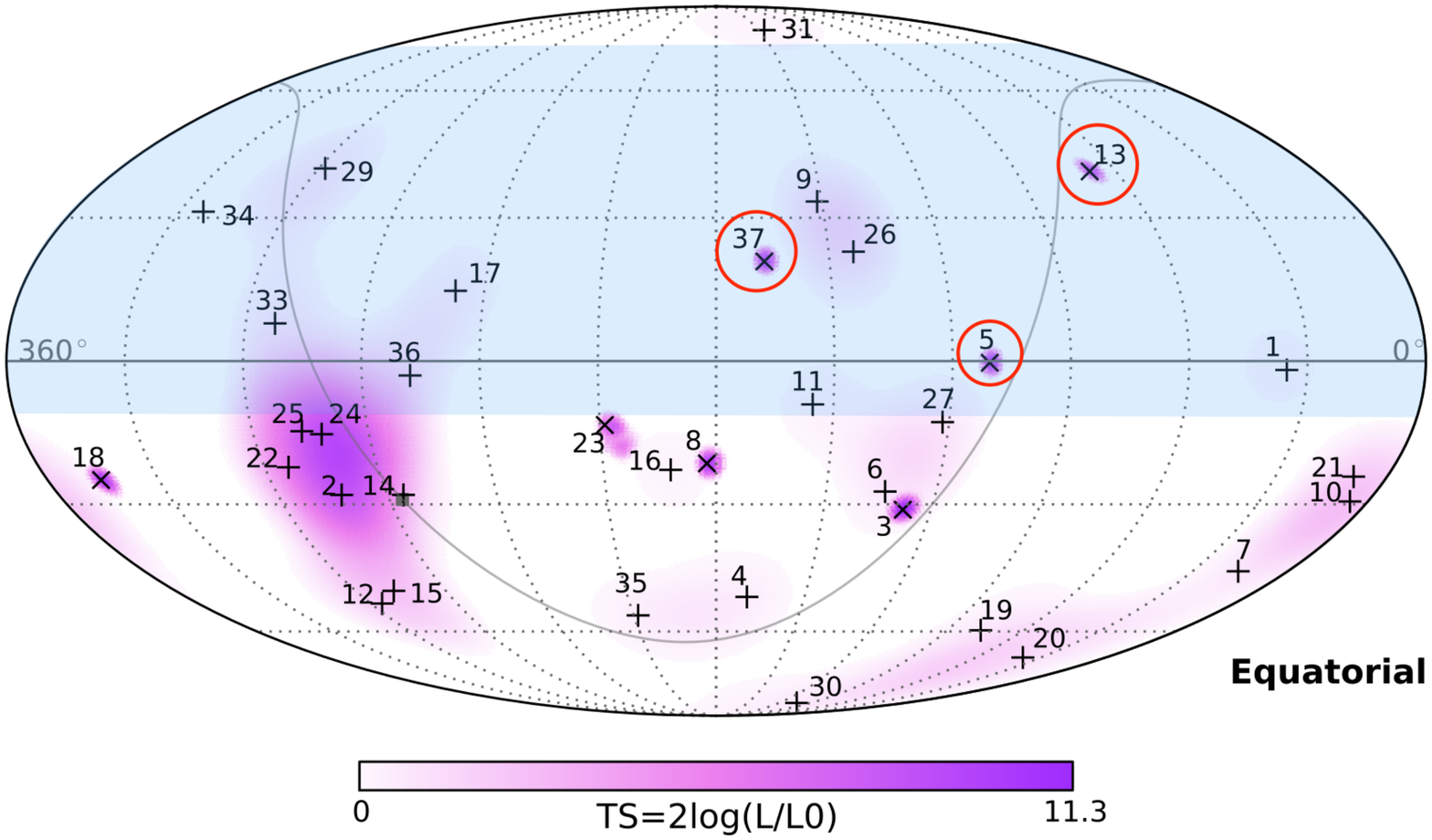} \hfill
\includegraphics[scale=0.31]{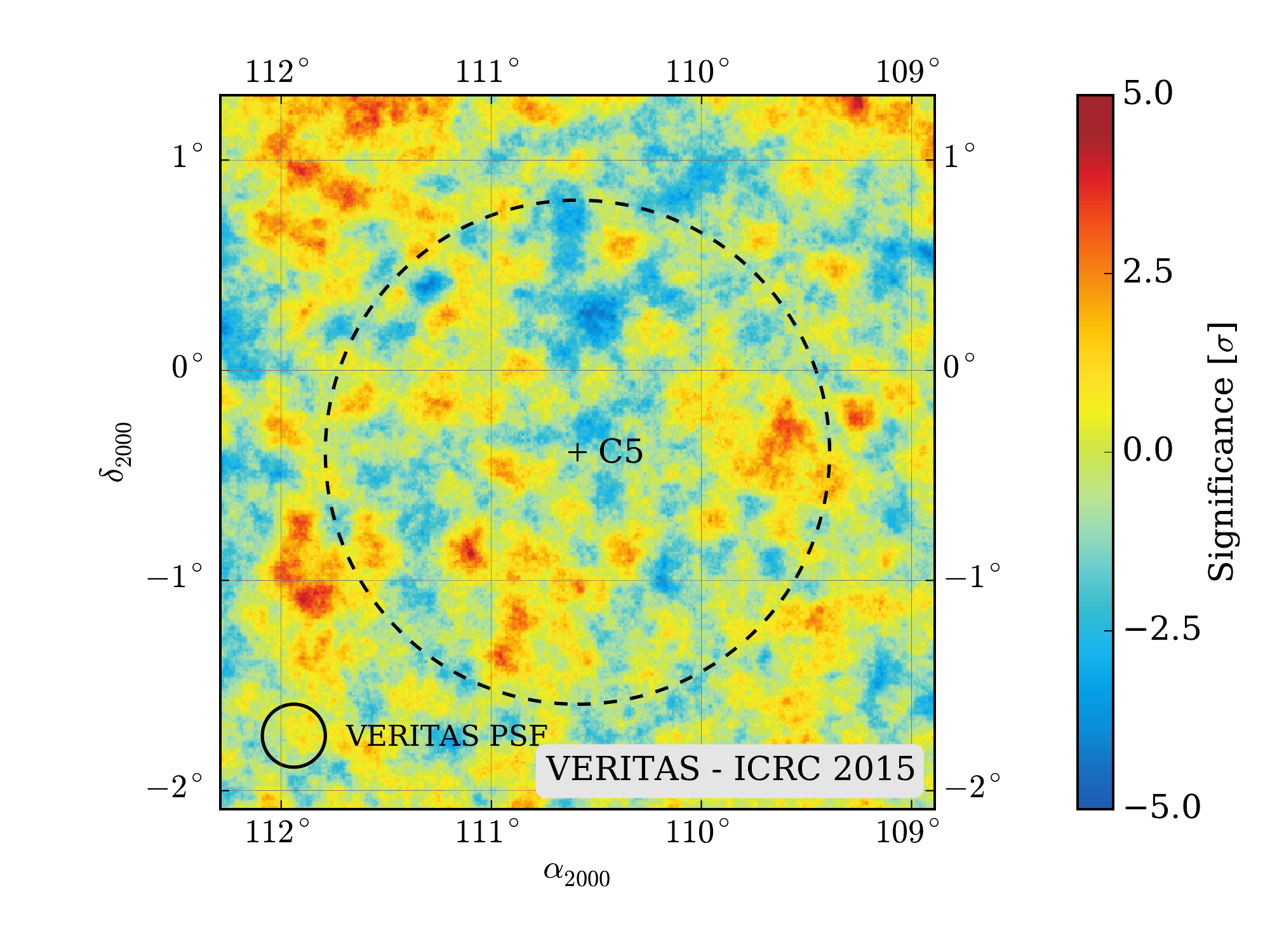} 
\end{center}
\caption{ Left: (a) IceCube sky map of contained high-energy neutrino events.The positions of the IceCube contained muon tracks (C5, C13, and C37) have been circled in red. Figure from~\cite{icecube_skymap}. Right: (b) Preliminary VERITAS skymap showing significance of detection at the position of C5. The dashed circle represents the angular uncertainty in the neutrino position as estimated by IceCube. Figure and caption from~\cite{icecube_veritas}.}  
\end{figure}

\section{Summary} 
This review presents some of the highlights from VERITAS in the last ten years. VERITAS recently marked ten years of successful operations with its full four-telescope array with a celebration and conference in Arizona in 2017 June (see the conference website \citep{veritas_10yr} for a set of talks highlighting the history and scientific accomplishments of VERITAS, as well as an overview from partner IACTs and multi-messenger observatories). The outlook in the field of very high energy astrophysics, with the upcoming Cherenkov Telescope Array  is excellent~\citep{cta_ksp}, and synergies with multi-messenger instruments such as IceCube, HAWC, and LIGO and VIRGO look promising. With its superior angular resolution, VERITAS is able to follow up on any new discoveries announced by multi-messenger partners in the Northern Hemisphere, with the potential of finding an electromagnetic counterpart in the  energy range 85 GeV to 30 TeV. The recent detection of gamma rays from the flaring blazar TXS 0506+056~\citep{magic_icecube, fermi_icecube}, in the field of the IceCube neutrino 170922A~\citep{ic170922}  was an exciting event, opening up new avenues in multi-messenger astronomy. The first-ever detections of electromagnetic counterparts to the LIGO event 20170817 has demonstrated the promise of multi-messenger astronomy~\citep{GBM170817} and new avenues for exploring the Universe have opened up using neutrinos and gravitational waves. 

Figure 18 shows the differential sensitivities of the different gamma-ray experiments currently in use and planned for the future~\citep{cta_obs}. The combined sensitivities from approximately 10 GeV to 100 TeV will allow deep studies to probe particle astrophysics processes in the cosmos. Future synergies will provide access to the high energy Universe using multiple messengers, not only with photons and cosmic rays, but also gravitational waves and neutrinos. 

\begin{figure}[h!]
\begin{center}
\includegraphics[scale=0.35]{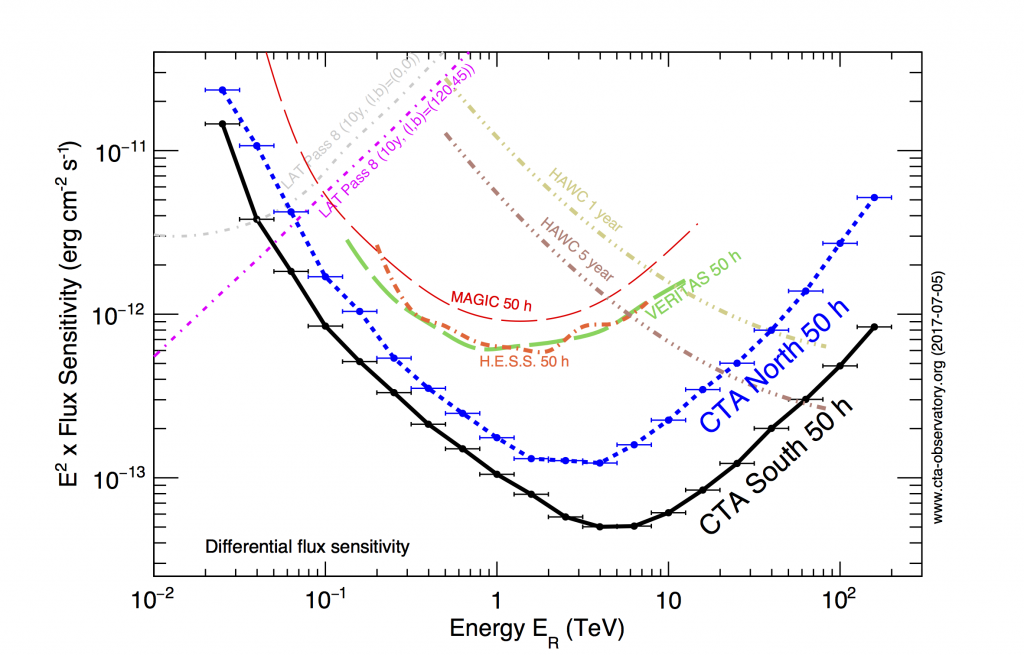} 
\end{center}
\caption{Differential sensitivities of the current generation of gamma-ray instruments. The curves for  and HAWC are scaled by a factor 1.2 to account for the different energy binning.The curves shown above give only an indicative comparison of the sensitivity of the different instruments, as the method of calculation and the criteria applied are different. A comparison with CTA is shown. Figure and caption from the CTA Observatory website~\cite{cta_obs}.}
\end{figure}

\bigskip
\begin{table}
\begin{small}
\begin{tabular}{|l|c|c|c|}
\hline
Source Name & Class &Redshift \\
{\bf(Extragalactic)}            &       &        $ z$     \\
\hline
Mrk 421                & HBL        & 0.031  \\
Mrk 501                & HBL        & 0.034  \\
1ES 2344+514    & HBL       & 0.044  \\
1ES 1959+650    & HBL       & 0.048  \\
1ES 1727+501    & HBL       & 0.055  \\
BL Lacertae         & LBL        & 0.069  \\
1ES 1741+196 & HBL & 0.084 \\
W Comae             & IBL         & 0.102  \\
RGB J0521.8+112 & IBL/HBL& 0.108 \\ 
RGB J0710+591 & HBL       & 0.125  \\
H 1426+428        & HBL        & 0.129  \\
B2 1215+303      & IBL/HBL & 0.131\\
S3 1227+25 & LBL & 0.135 \\
1ES 0806+524    & HBL       & 0.138  \\
1ES 0229+200    & HBL       & 0.140  \\
1ES 1440+122    & IBL/HBL& 0.163 \\
RX J0648.7+1516& HBL     & 0.179  \\
1ES 1218+304    & HBL       & 0.184  \\
RBS 0413             & HBL       & 0.190  \\
1ES 0647+250    & HBL & 0.203? \\
1ES 1011+496    & HBL       & 0.212 \\ 
MS 1221.8+2452 & HBL & 0.218 \\
1ES 0414+009    & HBL       & 0.287  \\
1ES 0502+675    & HBL       & 0.34\\
1ES 0647+250    & HBL       & $\sim 0.45$ \\
PG 1553+113      & HBL       & $0.43< z < 0.58$  \\
3C 66A                  & IBL        & $0.33<z<0.41$  \\
PKS 1222+216 & FSRQ & 0.432 \\
1ES 0033+595 & HBL & 0.467? \\
PKS 1424+240    & IBL/HBL& 0.604    \\
M87                       & FR I         & 0.0044  \\
M82                       & Starburst & 3.9 Mpc \\
HESS J1943+213 & HBL? & ? \\
RGB J2056+496 & HBL & ? \\
RGB J2243+203 & IBL/HBL & $>0.39$\\
PKS 1441+25 & FSRQ & 0.939 \\
\hline
\end{tabular}
\end{small}\\
\vfil
{\footnotesize Table 1: Extragalactic sources of TeV gamma-ray emission detected by VERITAS. Some redshifts are considered uncertain.}
\end{table}

\noindent
VERITAS research is supported by grants from the U.S. Department of Energy Office of Science, the U.S. National Science Foundation and the Smithsonian Institution, and by NSERC in Canada. The primary author acknowledges support from the US National Science Foundation through NSF grant PHYS-1505811. We acknowledge the excellent work of the technical support staff at the Fred Lawrence Whipple Observatory and at the collaborating institutions in the construction and operation of the instrument. The VERITAS Collaboration is grateful to Trevor Weekes for his seminal contributions and leadership in the field of VHE gamma-ray astrophysics, which made this study possible.

\end{document}